\documentclass[aps,pra,twocolumn,footinbib,superscriptaddress]{revtex4-2} 
	
	\usepackage{mathtools,amsmath}
	\usepackage{booktabs}
	\usepackage{graphicx} 
	\usepackage[breaklinks=true,colorlinks,citecolor=blue,linkcolor=blue,urlcolor=blue]{hyperref}
	\usepackage[notrig]{physics}
	\usepackage{bbold}
	\usepackage[scientific-notation=true]{siunitx}
	\usepackage[dvipsnames]{xcolor}
	
	\usepackage[normalem]{ulem}
	
	\renewcommand{\Re}{\mathop{\text{Re}}\nolimits}
	\renewcommand{\Im}{\mathop{\text{Im}}\nolimits}

	\renewcommand{\Tr}{\mathop{\text{Tr}}\nolimits}
	\definecolor{mygray}{gray}{0.6}

\begin{document}
\title{Identifying optimal cycles in quantum thermal machines with reinforcement-learning}

\author{Paolo A. Erdman}
\email{p.erdman@fu-berlin.de}
\affiliation{Freie Universit{\" a}t Berlin, Department of Mathematics and Computer Science, Arnimallee 6, 14195 Berlin, Germany}

\author{Frank No{\'e}}
\email{frank.noe@fu-berlin.de}
\affiliation{Freie Universit{\" a}t Berlin, Department of Mathematics and Computer Science, Arnimallee 6, 14195 Berlin, Germany}
\affiliation{Freie Universit{\" a}t Berlin, Department of Physics, Arnimallee 6, 14195 Berlin, Germany}
\affiliation{Rice University, Department of Chemistry, Houston, TX 77005, USA}
\affiliation{Microsoft Research, Cambridge, UK}

\begin{abstract}
The optimal control of open quantum systems is a challenging task but has a key role in improving existing quantum  information processing  technologies. We introduce a general framework based on Reinforcement Learning to discover optimal thermodynamic cycles that maximize the power of out-of-equilibrium quantum heat engines and refrigerators. We apply our method, based on the soft actor-critic algorithm, to three systems: a benchmark two-level system heat engine, where we find the optimal known cycle; an experimentally realistic refrigerator based on a superconducting qubit that generates coherence, where we find a non-intuitive control sequence that outperform previous cycles proposed in literature; a heat engine based on a quantum harmonic oscillator, where we find a cycle with an elaborate structure that outperforms the optimized Otto cycle. We then evaluate the corresponding efficiency at maximum power.
\end{abstract}

\maketitle

\section{Introduction}
Thermal machines convert between thermal and mechanical energy in a controlled manner. Examples include heat engines such as steam and Otto engines, that extract useful work from a temperature difference, and refrigerators, that extract heat from a cold bath.
Quantum thermal machines (QTMs) perform thermodynamic cycles via nanoscale systems that can be as small as single particles or two-level quantum systems (qubits). 
Quantum heat engines and refrigerators could find applications in heat management at the nanoscale \cite{fagas2014}, or for on-chip active cooling \cite{pekola2015,giazotto2006}.

Quantum thermodynamics is a rapidly growing research area that aims at the understanding, 
design and optimization of QTMs \cite{binder2019}. 
An open fundamental question is whether quantum effects can boost the performance of QTMs \cite{pekola2015,vinjanampathy2016,binder2019}. On the other hand, understanding how to optimally control the non-equilibrium dynamics of open quantum systems is a complicated task which can improve existing quantum information processing technologies. 

Nowadays it is possible to construct devices which behave as quantum systems with few degrees of freedom in platforms such as trapped ions \cite{friedenauer2008,blatt2012}, electron spins associated with nitrogen-vacancy centers \cite{childress2006}, circuit quantum electrodynamics \cite{wallraff2004}, and quantum dots \cite{peta2005}, and to control their state through time-dependent controls, such as electro-magnetic pulses or gate voltages.
The heat flow across these systems has been measured \cite{ronzani2018,dutta2019,senior2020,maillet2020}, and recent experimental realizations of QTMs have been reported \cite{rossnagel2016,josefsson2018,klatzow2019,lindenfels2019,maslennikov2019,peterson2019,prete2019,horne2020}.

While the laws of thermodynamics pose universal constraints on the efficiency of thermal machines, regardless of their classical or quantum nature, they do not pose any restriction on the dynamics of the system, thus on the speed at which it operates. Therefore, it is crucial to study the power to discover potential benefits of using QTMs. However, optimizing the power is a challenging task: having to operate in finite-time, the state can be driven far from equilibrium, requiring us to model the full dynamics of the quantum system. Furthermore, strategies are needed to identify optimally controlled cycles.

Power maximization of QTMs \cite{alicki1979,binder2019} is generally carried out either in specific regimes, or assuming \textit{a-priori} a specific shape of the control-cycle. Within the slow-driving \cite{esposito2010_prl,wang2011,avron2012,ludovico2016,cavina2017_prl,abiuso2018,bhandari2020} and fast-driving regime, general strategies have been recently derived \cite{abiuso2020_prl,abiuso2020_entropy,cavina2020}. 
Beyond these regimes, common strategies consider specific cycle structures \cite{arrachea2007,esposito2010_pre,juergens2013,campisi2015,dann2020,molitor2020}, such as the celebrated Otto cycle \cite{feldmann1996,feldmann2000,rezek2006,quan2007,abah2012,allahverdyan2013,zhang2014,campisi2016,karimi2016,kosloff2017,watanabe2017,deffner2018,gelbwaser2018,chen2019,pekola2019,das2020}, and optimize specific aspects of the cycle. Shortcuts to adiabaticity  \cite{berry2009,deng2013,torrontegui2013,campo2014,cakmak2018,deng2018,funo2019,villazon2019}, and variational optimization strategies have also been employed \cite{cavina2018,suri2018,menczel2019_prb}. The impact of quantum effects on the performance of QTMs is not straightforward. Several studies have found quantum advantages \cite{scully2011,uzdin2015,jaramillo2016,watanabe2017,deffner2018,das2020}, while coherence-induced power losses were proven in linear response for small driving amplitudes  \cite{brandner2017} and in specific models \cite{kosloff2002,rezek2006,karimi2016,cavina2018,pekola2019}.

In general, there is no guarantee that typical regimes and specific cycles considered in literature are optimal for power maximization.
Overcoming this limitation may allow us to unlock quantum advantages in power extraction. This calls for the development of powerful search strategies to tackle power-maximization without relying on specific control sequences or assumptions. 

In this manuscript we propose a Reinforcement Learning (RL) \cite{sutton2018} based approach to optimize the performance of QTMs. Specifically, we use a generalization of Soft Actor-Critic methods \cite{haarnoja2018_pmlr,haarnoja2018_arxiv_sac} for combined discrete and continuous actions, introduced in the context of robotics and videogames \cite{christodoulou2019,delalleau2019}, to discover thermodynamic cycles that deliver maximum power. 
RL has received a great deal of attention for its success at mastering complicated tasks beyond human-level such as playing video games \cite{mnih2015,vinyals2019}, the board-game Go \cite{silver2017} and for robotic applications \cite{haarnoja2018_arxiv_walk}. RL has been recently used for accurate quantum state preparation \cite{bukov2018,an2019,dalgaard2020,mackeprang2020}, outperforming previous state-of-the-art methods \cite{niu2019,zhang2019}, to minimize entropy production in closed quantum systems \cite{sgroi2021}, for fault-tolerant quantum computation \cite{sweke2021}, and machine learning methods have been used for quantum thermometry \cite{luiz2021}.

Our RL-based scheme for power maximization of QTMs is generic in that it makes no assumptions on the shape of the control cycle. Rather, it starts from scratch, allowing the RL agent to arbitrarily couple or decouple the quantum system from any bath, and to arbitrarily manipulate the control parameter.
We apply our approach to three paradigmatic systems that have been well studied in literature: 
(i) a benchmark heat engine based on a two-level system, where our approach automatically finds the known maximum power cycle \cite{erdman2019_njp}. 
(ii) an experimentally realistic refrigerator based on a superconducting qubit coupled to resonant circuits \cite{karimi2016} which generates coherence during its cycle. Our RL approach discovers a new and non-intuitive cycle that outperforms previous proposals \cite{karimi2016,pekola2019,funo2019}.  
(iii) a heat engine based on a harmonic oscillator \cite{rezek2006}, where we find a cycle with an elaborate structure that shares qualitative similarities with the Otto cycle, but which performs better thanks to additional features. We complement the study of these systems by analyzing the corresponding efficiency at maximum power.

The complexity and structure of these cycles demonstrates both the ability of the RL agent to choose actions based on a long-term return, and the complicated non-analytic nature of optimal cycles. Our results also show that the celebrated Otto cycle is not in general optimal for power maximization. Furthermore, we show that the detrimental effect of coherence on the performance of QTMs \cite{kosloff2002}, specifically observed both in the superconducting qubit and harmonic oscillator cases, can be mitigated thanks to carefully crafted cycles.

\section{Results}

\subsection{Quantum Thermal Machines}
\label{sec:qtm}

 \begin{figure}[!tb]
	\centering
	\includegraphics[width=0.99\columnwidth]{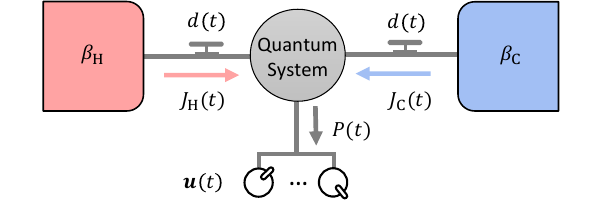}
	\caption{\textbf{Schematic representation of a quantum thermal machine.} A hot (cold) bath at inverse temperature $\beta_\mathrm{H}$ ($\beta_\mathrm{C}$), represented by the red (blue) box, can be coupled to the quantum system QS, gray circle, enabling a heat flux $J_\mathrm{H}(t)$ ($J_\mathrm{C}(t)$). The quantum system is controlled by an external agent through a set of control parameters $\mathbf{u}(t)$, which allow power exchange $P(t)$, and through a discrete control $d(t)=\{\mathrm{Hot}, \mathrm{Cold}, \mathrm{None}\}$ which determines which bath (if any) is coupled to the QS.}
	\label{fig:setup}
\end{figure}
We describe a QTM by a quantum system (QS), acting as a working medium, that can exchange heat with a hot (H) or cold (C) thermal bath characterized by inverse temperatures $\beta_\mathrm{H}<\beta_\mathrm{C}$ (Fig.~\ref{fig:setup}). We can control the evolution of the QS and exchange work with it though a set of time-dependent control parameters $\mathbf{u}(t)$. 
In classical thermal machines, the working medium could be a gas in a cylinder, and $\mathbf{u}(t)$ could be the time-dependent position of the piston which influences the state of the gas and allows us to exchange energy.
In a QTM, the working medium is a quantum system whose Hamiltonian can be parametrized by the control variables $\mathbf{u}(t)$ \cite{lekscha2018}.

Here we study finite-time thermodynamics of QTMs within the Markovian regime using the commonly employed master equation \cite{gorini1976,lindblad1976,breuer2002,yamaguchi2017}. This approach describes the time-evolution of the reduced density matrix of the QS, $\hat{\rho}(t)$, under the assumption of weak system-bath interaction. Setting $\hbar=1$, the master equation reads
\begin{equation}
	\frac{\partial }{\partial t} {\hat{\rho}}(t) = -i\left[ \hat{H}[\mathbf{u}(t)], \hat{\rho}(t)\right] + \sum\nolimits_{\alpha} \mathcal{D}^{(\alpha)}_{\mathbf{u}(t),d(t)}[\hat{\rho}(t)],
	\label{eq:lindblad}
\end{equation}
where $\hat{H}[\mathbf{u}(t)]$ is the Hamiltonian of the QS which depends explicitly on time via the control parameters $\mathbf{u}(t)$, $[\cdot,\cdot]$ denotes the commutator, and $\mathcal{D}^{(\alpha)}_{\mathbf{u}(t),d(t)}[\cdot]$ describes the effect of the coupling between the QS and bath $\alpha = \mathrm{H}, \mathrm{C}$. $d(t) = \{\mathrm{Hot}, \mathrm{Cold}, \mathrm{None}\}$ is an additional discrete control parameter which allows us to choose which bath (if any) is coupled to the  QS. 
We compute the extracted power $P(t)$ and the instantaneous heat flux $J_\alpha(t)$ flowing out of bath $\alpha$ in the standard way \cite{alicki1979} which guarantees the validity of the first law of thermodynamics $\partial{U}(t)/(\partial t) = -P(t) + \sum_\alpha J_\alpha(t)$, the internal energy being defined as $U = \mathrm{Tr}[\hat{\rho}(t)\hat{H}[\mathbf{u}(t)]]$ (see Methods for details).
The two main thermal machines we consider are the heat engine and the refrigerator. A heat engine is used to extract work, while a refrigerator is used to extract heat from the cold bath. Therefore, we define
\begin{eqnarray}
	P_{[\mathrm{E}]}(t) \equiv  \sum_{\alpha=\mathrm{H},\mathrm{C}} J_\alpha(t), &\quad
	P_{[\mathrm{R}]}(t) \equiv J_\mathrm{C}(t),
\end{eqnarray}
respectively as the instantaneous power of a heat engine E (since the total heat extracted coincides with the work if the internal energy difference is zero), and as the instantaneous cooling power of a refrigerator R. 

Our goal is to determine the optimal driving, i.e. to determine the functions $\mathbf{u}(t)$ and $d(t)$ that maximize the average power in the long run. We thus define the following exponentially weighted average of the power
\begin{equation}
	\langle P_{[\nu]}\rangle = \bar{\gamma} \int_0^\infty e^{-\bar{\gamma} t} P_{[\nu]}(t)\,dt,
	\label{eq:avg_p_def}
\end{equation}
where $\bar{\gamma}$ determines the timescale over which we average. 
While we do not enforce any periodic structure on the controls $\mathbf{u}(t)$ and $d(t)$, we expect the RL agent to automatically discover the optimality of periodically driving the QTM and the corresponding driving period. 
The intuition is the following: in the short term, we can maximize the power by taking advantage of the state preparation of the system, for example by extracting all the free energy from the system. However, the amount of work that can be extracted this way is bounded, while the work that can be extracted through cycles scales with the number of performed cycles, i.e. with time. Therefore in the long run, i.e. for small enough $\bar{\gamma}$, we expect the maximization of Eq.~(\ref{eq:avg_p_def}) to naturally discover thermodynamic cycles and to prevent the exploitation of transient effects. We confirm this hypothesis in all QTMs studied below.

\subsection{Reinforcement Learning for Quantum Thermal Machines}
 \begin{figure}[!tb]
	\centering
	\includegraphics[width=0.99\columnwidth]{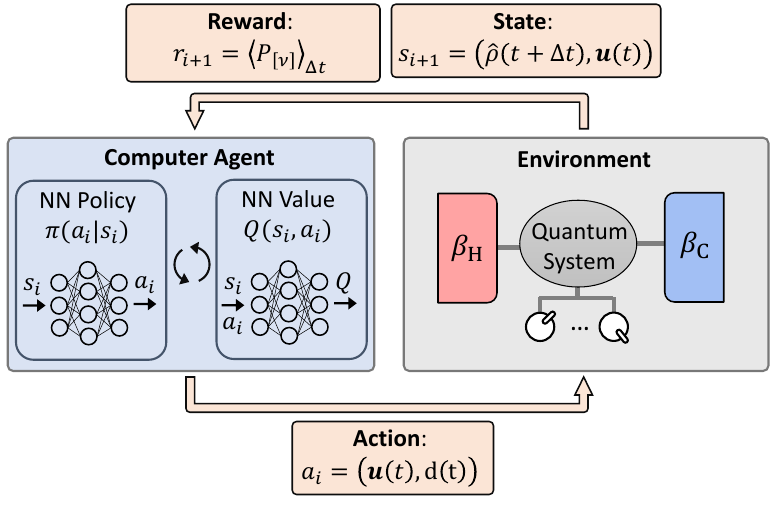}
	\caption{\textbf{Schematic representation of the learning process.} A computer agent (blue box) chooses an action $a_i$ at time step $i$ based on the current state $s_i$ of the environment (gray box) through the policy function $\pi(a_i|s_i)$. The action, which encodes the control ($\mathbf{u}(t)$, $d(t)$), is passed to the environment (lower  arrow) which evolves the quantum state $s_i$ of the machine based on the action $a_i$, and computes the average power during the time step as reward. The new state $s_{i+1}$ and reward $r_{i+1}$ are returned to the agent (upper arrow), which uses this information to improve $\pi(a|s)$ using the soft actor-critic algorithm, which learns also the value function $Q(s,a)$. Both $\pi(a|s)$ and $Q(s,a)$ are parameterized using fully-connected neural networks (NNs). This process is reiterated until convergence of the policy.}
	\label{fig:rl}
\end{figure}
We formulate the power optimization problem as a discounted, continuing RL task. As we will demonstrate, this approach is able to learn far-from-equilibrium finite-time thermodynamic cycles with high performance.

RL is a general framework that tackles optimization problems formulated in the following way. As schematically shown in Fig.~\ref{fig:rl}, a \textit{computer agent} (blue box) must learn to master some task by repeated interactions with some \textit{environment} (gray box). Discretizing time in time steps $\Delta t$, we denote with $s_i \in\mathcal{S}$ the state of the environment at time $t=i \Delta t$, where $\mathcal{S}$ is the \textit{state space}. The agent must choose an action $a_i\in \mathcal{A}$ to perform on the environment based on its current \textit{policy} (lower orange arrow). $\mathcal{A}$ is the \textit{action space}, and the policy $\pi(a_i|s_i)$ is a function that describes the probability distribution of choosing action $a_i$, given that the environment is in state $s_i$. The environment then evolves its state according to the chosen action, and provides feedback back to the agent by returning the updated state $s_{i+1}$ and a scalar quantity $r_{i+1}$ known as the \textit{reward} (upper orange arrow). 
This procedure is reiterated for a large number of time steps. 

At every time step $t_i$, the aim of the agent is to use the feedback it receives from the environment to learn an optimal policy that maximizes, in expectation, the \textit{return}, i.e.  the total future reward it receives from the environment, defined as
\begin{equation}
	r_{i+1} + \gamma r_{i+2} + \gamma^2 r_{i+3} + \dots = \sum_{k=0}^\infty \gamma^k r_{i+1+k},
	\label{eq:return_def}
\end{equation}
where $\gamma \in [0,1)$ is the \textit{discount factor} which determines how much we are interested in future rewards, as opposed to immediate rewards.

We now turn to the power maximization problem. Discretizing time in steps $\Delta t$, we search for protocols $\mathbf{u}(t)$ and $d(t)$ that are constant during each time step. As shown in Fig.~\ref{fig:rl}, we choose as action space $\mathcal{A} = \{ (\mathbf{u}, d) \,|\, \mathbf{u}\in \mathcal{U}, d\in\{ \mathrm{Hot}, \mathrm{Cold}, \mathrm{None} \} \}$, where $\mathcal{U}$ is the continuous set of accessible controls, which can account for any experimental limitation, and $d$ is the discrete action, motivated by typical thermodynamic cycles,  which determines which bath (if any) is coupled to the QS.
We choose the physical quantum states of the QS and the last chosen action as state space, i.e.  $\mathcal{S} = \{ (\hat{\rho}, \mathbf{u})\,|\, \hat{\rho}\in \mathcal{D}, \mathbf{u}\in \mathcal{U} \}$, where $\mathcal{D} = \{\hat{\rho} \,|\, \hat{\rho}\geq 0, \mathrm{Tr}[\hat{\rho}]=1\}$ is the space of density matrices. 
Crucially, we choose as reward 
\begin{equation}
	r_{i+1} = \langle P_{[\nu]} \rangle_{\Delta t} \equiv \frac{1}{\Delta t} \int_t^{t+\Delta t} P_{[\nu]}(\tau)\, d\tau,
	\label{eq:r_qtm}
\end{equation}
which is the average power of the machine during the time interval $[t,t+\Delta t]$. Plugging  Eq.~(\ref{eq:r_qtm}) into Eq.~(\ref{eq:return_def}), we see that the aim of the agent is to maximize the average power $\langle P_{[\nu]} \rangle$ introduced in Eq.~(\ref{eq:avg_p_def}), where $\bar{\gamma} = -\ln\gamma/\Delta t$ (see Methods for details).
In the RL notation, $\gamma$ sets the timescale for the power averaging, with $\gamma \to 1$ corresponding to long term averaging.

Our agent has no prior knowledge of quantum dynamics, nor of thermodynamic cycles: the evolution of the state from $s_i = \hat{\rho}(t)$ to $s_{i+1}=\hat{\rho}(t+\Delta t)$ and the computation of the rewards $r_i$ is performed by the environment.
We learn the optimal policy employing the soft actor-critic method, which relies on learning also a value function $Q(s,a)$, generalized to a combination of discrete and continuous actions \cite{haarnoja2018_pmlr,haarnoja2018_arxiv_sac,christodoulou2019,delalleau2019}. In this approach, the policy function $\pi(a|s)$ plays the role of an ``actor'' that chooses the actions to perform, while the value function $Q(s,a)$ plays the role of a ``critic'' that judges the choices made by the actor, thus providing feedback to improve the actor's behavior.  Both $\pi(a|s)$ and $Q(s,a)$ are parameterized using fully-connected neural networks with two hidden layers, and they are determined by minimizing the loss functions in Eqs.~(\ref{eq:q_loss}) and (\ref{eq:pi_loss}) in Methods using the ADAM optimization algorithm \cite{kingma2014}. The gradient of the loss functions is computed off-policy, over a batch of past experience which is recorded and stored in a replay buffer, using backpropagation (see Methods for details).

\subsection{Case Studies}
\label{eq:case_studies}
In this section we prove the validity of our RL-based approach by applying it to three different systems, namely a heat engine based on a two-level system, a refrigerator based on a superconducting qubit, and a heat engine based on a quantum harmonic oscillator. While the results presented below were obtained performing a single training, in Methods we show that our RL approach reliably converges to solutions with nearly the same performance across multiple trainings.

\paragraph{Two-level system heat engine.}
We first benchmark our method on a two-level system for which the optimal control cycle is known: 
\begin{equation}
	\hat{H}[u(t)] = \frac{E_0 u(t)}{2}\,\hat{\sigma}_z,
\end{equation}
where $u(t)$, which determines the energy gap of the two-level system, is our single control parameter, $E_0$ is a fixed energy scale and $\hat{\sigma}_z$ is a Pauli matrix. We consider the qubit to be coupled to fermionic baths with flat density of states (for example, a single-level quantum dot tunnel-coupled to metallic leads), with thermalization timescales fixed by the rates $\Gamma_\alpha$ (see Methods for details).

 \begin{figure}[!tb]
	\centering
	\includegraphics[width=0.99\columnwidth]{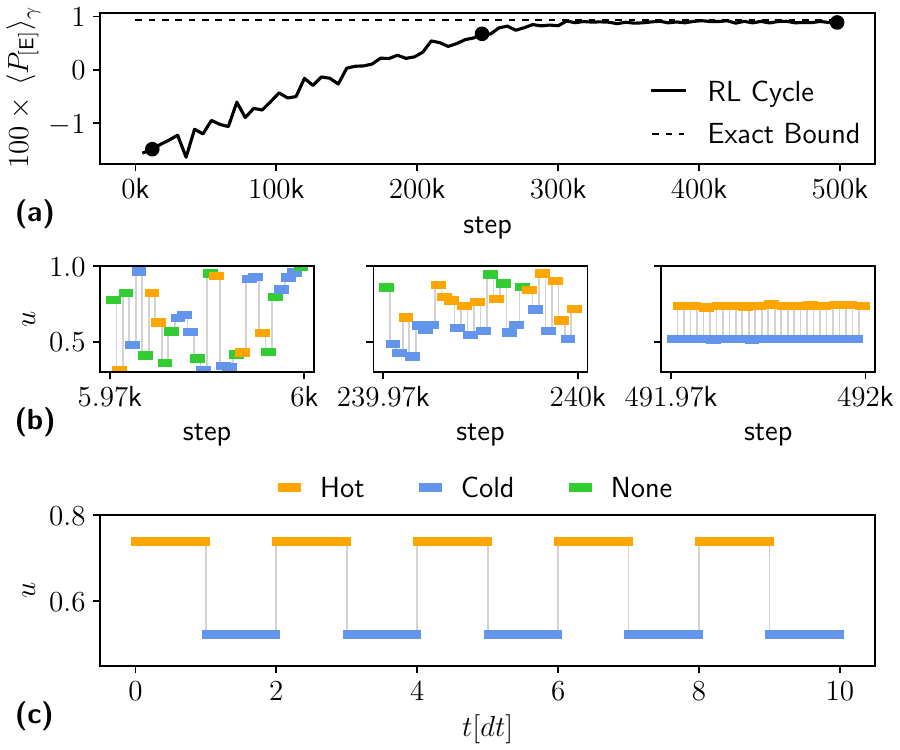}
    \caption{\textbf{Results of training the agent on the two-level heat engine model.} (a): Running average of the extracted power $\langle P_{[\mathrm{E}]}\rangle_\gamma$ as a function of the step during the whole training. The dashed line represents the theoretical upper bound derived in Ref.~\cite{erdman2019_njp}. (b): Actions chosen by the agent, represented by the value of $u$, as a function of step, zoomed around the three black circles in panel (a). The color represents action $d$, i.e. the bath coupled to the QS. (c): Final cycle found by the agent as a function of time. The parameters used for training are: $\Gamma_\mathrm{H}=\Gamma_\mathrm{C}=1$, $\beta_\mathrm{H}=1$, $\beta_\mathrm{C}=2$, $\mathcal{U}= [0.3,1]$, $E_0=2.5$, $\Delta t=0.5$, and $\gamma=0.995$. }
	\label{fig:qubit_engine}
\end{figure}
Our results are shown in Fig.~\ref{fig:qubit_engine}.
Fig.~\ref{fig:qubit_engine}a shows the
running average of the power $\langle P_{[\mathrm{E}]}\rangle_\gamma(i)\equiv (1-\gamma)\sum_{k=0}^{i}\gamma^k r_{i-k}$, i.e. an exponentially weighed average of the past rewards with weight $\gamma$, as a function of the time steps. 
Fig.~\ref{fig:qubit_engine}b show the actions chosen by the agent, as a function of the steps, at three different moments during training highlighted by the black dots in Fig.~\ref{fig:qubit_engine}a. The position of the segments corresponds to the chosen value of $u$, while the color represents the discrete action $d$ (see legend).
The optimal cycle learned at the end of the training is shown in  Fig.~\ref{fig:qubit_engine}c.
Initially, the agent has no knowledge of the system, and its actions appear random (Fig.~\ref{fig:qubit_engine}b, left), producing negative power $\langle P_{[E]}\rangle_\gamma$ (Fig.~\ref{fig:qubit_engine}a). 
As expected, by performing random actions the agent is dissipating work into the heat baths, rather than extracting work. 
With increasing time, the agent gains experience and learns how to control the heat engine: $\langle P_{[E]}\rangle_\gamma$ increases, and structure appears in the chosen actions (Fig.~\ref{fig:qubit_engine}b, center and right). Eventually the policy converges, and $\langle P_{[E]}\rangle_\gamma$ saturates to a finite positive value. 

The optimal cycle for this model was derived in Ref.~\cite{erdman2019_njp}, and it corresponds to the exact structure discovered by the agent, i.e. a square wave alternating between the hot and the cold bath as fast as possible without spending any time disconnected from the baths. The black dashed line in Fig.~\ref{fig:qubit_engine}a shows the corresponding power. Notably, although the frequency of the learned cycle is limited by the choice of $\Delta t$, and although the values of $u(t)$ found by the agent are slightly different respect to the ones predicted by Ref.~\cite{erdman2019_njp} (the difference is $\approx 0.05$), the power it generates nearly coincides with the upper bound. This occurs because there is a manifold of near-optimal solutions.

We conclude the analysis of the two-level system heat engine by computing the \textit{efficiency at maximum power} (EMP), i.e. the thermodynamic efficiency of the heat engine, defined as the ratio between the extracted work and the input heat, while the engine is operated at maximum power. We compare the EMP both to the Carnot efficiency $\eta_\mathrm{c}=1-\beta_\mathrm{H}/\beta_\mathrm{C}$ and to the Curzon-Ahlborn efficiency $\eta_\mathrm{ca}=1-\sqrt{\beta_\mathrm{H}/\beta_\mathrm{C}}$. Despite not being a fundamental upper bound, the latter has received considerable attention in the literature for its simplicity, for being an upper bound to the EMP in various specific models \cite{curzon1975,schmiedl2007,esposito2010_prl,cavina2017_prl}, and it has been derived by general arguments from linear irreversible thermodynamics \cite{broeck2005}.
Interestingly, we find that the optimal cycle shown in Fig.~\ref{fig:qubit_engine}c delivers a large EMP corresponding to $100\%$ of $\eta_\mathrm{ca}$, equivalent to $59\%$ of $\eta_\mathrm{c}$.

\paragraph{Superconducting qubit refrigerator.}
We now consider a refrigerator based on an experimentally realistic system, i.e. a superconducting qubit coupled to two resonant circuits which behave as heat baths \cite{karimi2016}. As opposed to the previous setup, here coherence between the instantaneous eigenstates is generated while driving the system, since $[\hat{H}(u_1), \hat{H}(u_2)]\neq 0$ for $u_1\neq u_2$. This quantum effect was found to deter the power of this specific setup \cite{karimi2016,pekola2019,funo2019} and of arbitrary systems in linear response and for small driving amplitudes \cite{brandner2017}. 

As shown in Refs.~\cite{karimi2016,pekola2019,funo2019}, the system Hamiltonian is given by
\begin{equation}
	\hat{H}[u(t)] = - E_0\left[\Delta \hat{\sigma}_x + u(t)\hat{\sigma}_z  \right],
	\label{eq:h_fridge}
\end{equation}
where $E_0$ is a fixed energy scale, $\Delta$ characterizes the minimum gap of the system, and $u(t)$ is our control parameter. In this setup the coupling to the bath is fixed, and cannot be controlled. However, the qubit is resonantly coupled to the baths at different energies. 
The $u$-dependent coupling strength to the  C (H) bath is described by the function $\gamma^{(\text{C})}_u$ ($\gamma^{(\text{H})}_u$) that, as in Ref.~\cite{funo2019}, is peaked at $u=0$ ($u=1/2$) with a resonance width determined by the quality factor $Q_\text{C}$ ($Q_\text{H}$) (see Methods for details).

 \begin{figure}[!tb]
	\centering
	\includegraphics[width=0.99\columnwidth]{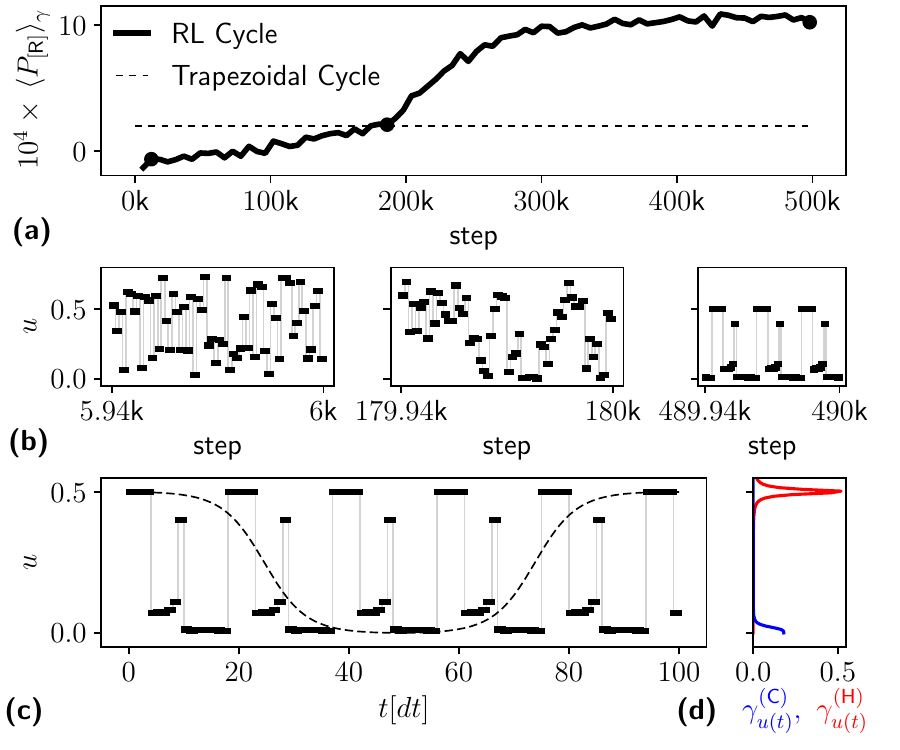}
\caption{\textbf{Results of training the agent on the superconducting qubit refrigerator model.} (a): Running average of the cooling power $\langle P_{[\mathrm{R}]}\rangle_\gamma$ as a function of the step during the whole training. The dashed line represents the maximum cooling power found in Ref.~\cite{funo2019} by optimizing a smoothed trapezoidal cycle. (b): Actions chosen by the agent, represented by the value of $u$, as a function of step, zoomed around the three black circles in panel (a). (c): Final deterministic cycle found by the agent (thick discontinuous line) and smoothed trapezoidal cycle (thin dashed line) whose power is given by the dashed line in panel (a), as a function of time. (d): coupling strength $\gamma^{(C)}_u$ (blue curve) and $\gamma^{(H)}_u$ (red curve) as a function of $u$ (on the y-axis). The parameters used for training are $g_\mathrm{H}=g_\mathrm{C}=1$, $\beta_\mathrm{H}=10/3$, $\beta_\mathrm{C} = 2\beta_\mathrm{H}$, $Q_\mathrm{H}=Q_\mathrm{C}=30$, $E_0=1$, $\Delta =0.12$, $\omega_\mathrm{H} = 1.03$, $\omega_\mathrm{C}=0.24$, $\mathcal{U}=[0,0.75]$, $\Delta t=0.98$, and $\gamma=0.995$.}
	\label{fig:qubit_fridge}
\end{figure}

Panels (a), (b) and (c) of Fig.~\ref{fig:qubit_fridge} report the results in the same style as in Fig.~\ref{fig:qubit_engine}, with the exception that all actions are black since there is no discrete choice to make, while Fig.~\ref{fig:qubit_fridge}d shows the coupling strength $\gamma^{(\text{C})}_u$ (blue curve) and $\gamma^{(\text{H})}_u$ (red curve) as a function of $u$ (on the y-axis).
 The parameters were chosen as in Fig.~7 of Ref.~\cite{funo2019}. As previously, the agent begins with random actions in the first steps, and the corresponding running average cooling power $\langle P_{[\mathrm{R}]}\rangle_\gamma$ is slightly negative, indicating that we are dissipating heat into the cold bath. As the agent gains experience, $\langle P_{[\mathrm{R}]} \rangle_\gamma$ increases until it saturates to a final positive value obtained with the cycle shown in Fig. \ref{fig:qubit_fridge}c (thick lines). 

Interestingly, this setup was partially optimized in Ref.~\cite{funo2019}. In their analysis, the authors fix a smoothed trapezoidal cycle $u(t)$ (Fig.~\ref{fig:qubit_fridge}c, dashed line) which was shown to outperform a sine and a trapezoidal cycle \cite{karimi2016}. They find that the cooling power is positive for large cycle periods $T$, and tends to zero as $T\to \infty$. However, the cooling power becomes negative as $T\to 0$ because the detrimental effect of the generation of coherence increases with the speed of the cycle. As a consequence, there is an intermediate optimal value of $T$. They find that $\langle P_{[\mathrm{R}]}\rangle \approx 2.3 \times 10^{-4}$ at this optimal choice (dashed line in Fig.~\ref{fig:qubit_fridge}a,c). Notably, our RL agent discovers a protocol with $\langle P_{[\mathrm{R}]}\rangle\approx 10.8 \times 10^{-4}$ using the same system parameters. This improvement is due to the non-intuitive additional step visible in Fig.~\ref{fig:qubit_fridge}c. Indeed, both the trapezoidal and the agent's cycle spend time in resonance with the baths at $u=0$ and $u=0.5$. However, the agent identifies also a third point, around $u\approx 0.13$, where it spends $\approx 1/4$ of the total cycle time and where the system is essentially decoupled from the bath, thus undergoing unitary evolution (as can be seen by the small values of $\gamma^{(\alpha)}_u$ corresponding to  $u\approx 0.13$ in Fig.~\ref{fig:qubit_fridge}d). 
This additional feature allows us to roughly extract the same amount of heat per cycle, but $5$ times faster. Interestingly, we verified that the trapezoidal cycle running at the same speed as the cycle found by the RL agent would yield negative power. As argued in \cite{karimi2016,pekola2019,funo2019}, this power loss is attributed to the generation of coherence during the cycle, so we can interpret the power enhancement achieved by our cycle as a mitigation of such detrimental effect. To confirm this, we rigorously quantify the generation of coherence in both cycles by computing the time-average of the \textit{relative entropy of coherence} \cite{baumgratz2014}. Indeed, we find that the trapezoidal cycle operated at the same speed as the cycle found by the RL agent generates almost twice as much coherence (see Methods for details). 

We conclude the study of the qubit-based refrigerator by evaluating the \textit{coefficient of performance} (COP) at maximum power, i.e. the ratio between the heat extracted from the cold bath and the input work, while the system is operated at maximum cooling power. We find that our cycle shown in Fig.~\ref{fig:qubit_fridge}c as a thick black line delivers a COP at maximum power that is $6\%$ of $C_\mathrm{c} = \beta_\mathrm{C}^{-1}/(\beta_\mathrm{H}^{-1} - \beta_\mathrm{C}^{-1})$, which is Carnot's upper bound to the COP. While this may appear as a rather low value, we notice that the COP at maximum power of a two-level system coupled to Fermionic of Bosonic baths is zero \cite{erdman2019_njp}, and that for on-chip cooling applications, the aim is typically to maximize the cooling power regardless of the efficiency of such a process.

\paragraph{Harmonic oscillator heat engine.}
At last, we consider a heat engine based on a collection of non-interacting particles confined in a harmonic potential \cite{rezek2006}. The Hamiltonian is given by
\begin{equation}
	\hat{H}[u(t)] = \frac{1}{2m} \hat{p}^2 + \frac{1}{2}m (u(t)w_0)^2 \hat{q}^2,
\end{equation}
where $m$ is the mass of the system, $w_0$ is a fixed frequency and $\hat{p}$ and $\hat{q}$ are the momentum and position operators. The control parameter $u(t)$ allows us to change the frequency of the oscillator. As in the qubit heat engine case, we let the agent choose which bath (if any) to couple to the oscillator. Since $[\hat{H}(u_1), \hat{H}(u_2)]\neq 0$ for $u_1\neq u_2$, also this system exhibits a power loss at finite driving speed ~\cite{kosloff2002,rezek2006}. The coupling to the baths, characterized by the thermalization rates $\Gamma_\alpha$, is modeled as in Ref.~\cite{rezek2006} (see Methods for details).

 \begin{figure}[!tb]
	\centering
	\includegraphics[width=0.99\columnwidth]{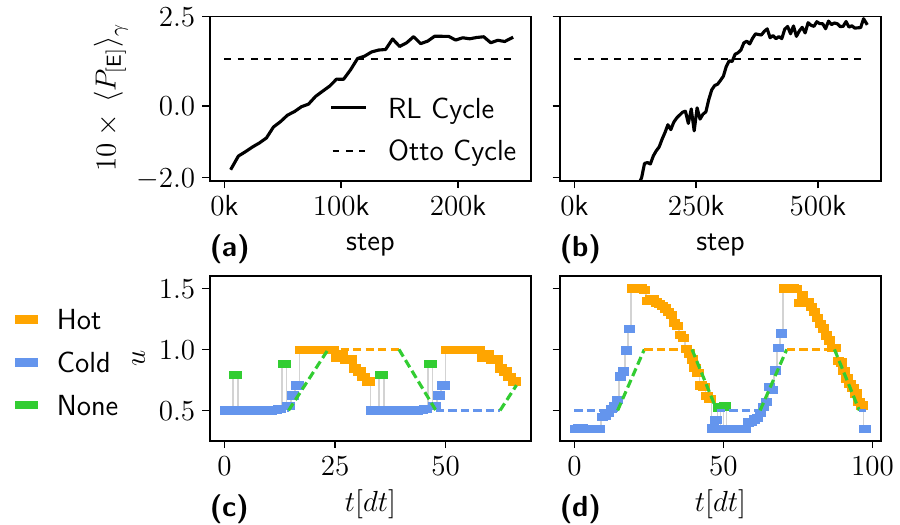}
	\caption{\textbf{Results of training the agent on the quantum harmonic oscillator heat engine model.} (a-b): Running average of the extracted power $\langle P_{[\mathrm{E}]}\rangle_\gamma$ as a function of the step during the whole training. The dashed line represents the maximum power found in Ref.~\cite{rezek2006} optimizing the duration of each stroke of an Otto cycle (see Methods for details). (c-d): Actions chosen by the agent, represented by the value of $u$, as a function of time. The color represents action $d$, i.e. the bath coupled to S. The discontinuous thick line is the final cycle discovered by the agent, while the thin dashed line is the Otto cycle whose power is given by the dashed line in panels (a-b). The parameters used for training are $\Gamma^{(\mathrm{H})}=\Gamma^{(\mathrm{C})}=0.6$, $\beta_\mathrm{H}=0.2$, $\beta_\mathrm{C} = 2$, $w_0=2$, $\Delta t=0.2$, and $\gamma=0.998$. In panels (a-c) we set $\mathcal{U}=[0.5,1]$ to enable a fair comparison with the Otto cycle of Ref.~\cite{rezek2006}, while in (b-d) we set a larger control range, i.e. $\mathcal{U}=[0.35,1.5]$. }
	\label{fig:harmonic_engine}
\end{figure}
The solid line in Figs.~\ref{fig:harmonic_engine}a and \ref{fig:harmonic_engine}b shows $\langle P_{[\mathrm{E}]}\rangle_\gamma$, as a function of the step, for the same system parameters (chosen as in the upper panel of Fig. 6 of Ref.~\cite{rezek2006}), but setting respectively $\mathcal{U}=[0.5,1]$ and $\mathcal{U}=[0.35,1.5]$. In both cases, the power is negative for small steps, while it converges to a positive value as the agent gains experience. The corresponding final cycles learned by the agent are shown in Figs.~\ref{fig:harmonic_engine}c and \ref{fig:harmonic_engine}d as thick lines.  These cycles display a quite elaborate structure which demonstrates the ability of the agent to perform planning for a long-time reward. Indeed, while the system is in contact with the cold bath, which happens roughly for $20$ time steps, energy is flowing into the bath, thus producing negative rewards. Nonetheless, the agent discovers that this step, required by the second law of thermodynamics, is necessary for power extraction in the long run.

We now compare the cycle discovered by the agent to the well-known Otto cycle often considered in literature. The authors of Ref.~\cite{rezek2006} study the power of this system by optimizing an Otto cycle, i.e. the cycle shown in Figs.~\ref{fig:harmonic_engine}c and \ref{fig:harmonic_engine}d as a dashed line. The authors then fix the value of the control $u_\mathrm{H}=1$ ($u_\mathrm{C}=0.5$) while in contact with bath H (C), and maximize the power by tuning the duration of each of the 4 segments composing the cycle. The resulting optimized Otto cycle is the one shown as a dashed line in Figs.~\ref{fig:harmonic_engine}c and \ref{fig:harmonic_engine}d, with corresponding power shown in Figs.~\ref{fig:harmonic_engine}a and \ref{fig:harmonic_engine}b as a dashed line. As we can see, even constraining the choice of $u(t) \in [u_\mathrm{C},u_\mathrm{H}]=[0.5,1]$ to the values chosen in Ref.~\cite{rezek2006}, the agent discovers a cycle that outperforms the optimized Otto cycle (Fig.~\ref{fig:harmonic_engine}a). If we further allow the agent to modulate $u(t)$ in a larger interval, we find a cycle with an even larger power (Fig.~\ref{fig:harmonic_engine}b).

The protocol discovered by the agent in Fig.~\ref{fig:harmonic_engine}a resembles the Otto cycle (both are flat for some time at $u=u_\mathrm{C}$ and $u=u_\mathrm{H}$), yet there are additional features. First, the agent ramps up the control $u(t)$ between $u_\mathrm{C}$ and $u_\mathrm{H}$ in a non-linear fashion while in contact with a bath, rather than decoupling the system. Next, there are two strong discontinuities where the system is abruptly disconnected from the baths for a short time (Fig.~\ref{fig:harmonic_engine}c, green segments). These additional features turn out to enhance the power.
At last, we notice that the cycle found by the agent in Fig.~\ref{fig:harmonic_engine}d is more regular than the one in Fig.~\ref{fig:harmonic_engine}c, and it further deviates from an Otto cycle. The crucial difference with respect to the Otto cycle therefore seems to be the ramping up of $u(t)$ while in contact with the baths, which probably benefits from simultaneously exchanging heat and modulating $u(t)$, rather than doing them in two separate strokes.

As in the two-level case, we evaluate the EMP of the discovered cycles shown in Fig.~\ref{fig:harmonic_engine}c and \ref{fig:harmonic_engine}d. In both cases the EMP is considerably high, corresponding respectively to $60\%$ and $78\%$ of $\eta_\mathrm{ca}$, equivalent to $46\%$ and $59\%$ of $\eta_\mathrm{c}$. Interestingly, the cycle shown in Fig.~\ref{fig:harmonic_engine}d yields both a higher power and a larger EMP than the cycle displayed in Fig.~\ref{fig:harmonic_engine}c.

\section{Discussion}
We introduced a general framework based on Reinforcement Learning to discover thermodynamic cycles that maximize the power of out-of-equilibrium quantum thermal machines, paving the way for a more systematic use of machine learning in the field of quantum thermodynamics. Using state-of-the-art machine learning techniques, we applied our method to three different paradigmatic setups. Our method found the optimal known solution in the benchmark system, while in the other systems it discovered new unintuitive and elaborate cycles which outperform previously proposed cycles. 

Our results show that the celebrated Otto cycle is not in general optimal for power extraction, and that carefully crafted cycles can mitigate coherence-induced power losses \cite{kosloff2002,karimi2016,brandner2017,pekola2019} without introducing additional controls as required by Shortcuts to Adiabaticity \cite{berry2009,deng2013,torrontegui2013,campo2014,cakmak2018,deng2018,funo2019,villazon2019}. 
As opposed to other optimal control techniques, such as the Pontryaghin Minimum Principle \cite{cavina2018,suri2018,menczel2019_prb}, our RL-based approach has the following advantages: it does not require any analytic calculation; it can handle both continuous and discrete controls (such as the choice of the heat bath), and it can naturally find discontinuous and irregular protocols; it can be applied as-is to arbitrarily complicated setups and it could be used to find optimal protocols also in the presence of noise in the controls.

Future research directions include applications to multi-particle systems, where many-body advantages might be revealed, and a systematic study of the mitigation of coherence-induced power losses. Interesting extensions of our framework include investigating the strong system-bath coupling regime going beyond a master equation approach \cite{gallego2014,gelbwaser2015,perarnau2018}, optimizing additional thermodynamic quantities, such as minimizing the fluctuations in the power output, and developing of a scheme that can be applied directly to experimental setups that does not require the knowledge of the quantum state.

\section{Methods}

\paragraph*{Physical model.}  
As discussed in the main text, we assume that the state evolves according to the Markovian master Eq.~(\ref{eq:lindblad}), which can be derived, also for non-adiabatic drivings \cite{yamaguchi2017}, in the weak system-bath coupling regime performing the usual Born-Markov and secular approximation \cite{gorini1976,lindblad1976,breuer2002} and neglecting the Lamb-shift contribution. We notice that since the RL agent produces piece-wise constant protocols, we are not impacted by possible inaccuracies of the master equation subject to fast parameter driving \cite{dann2018}, provided that $\Delta t$ is not smaller than the bath timescale. 
Without loss of generality, the dissipators can be expressed as \cite{lindblad1976,breuer2002}
\begin{multline}
	\mathcal{D}^{(\alpha)}_{\mathbf{u}(t),d(t)}[\hat{\rho}] = \lambda_\alpha[d(t)] \sum_k \gamma^{(\alpha)}_{k,\mathbf{u}(t)} \times \\
	( \hat{A}^{(\alpha)}_{k,\mathbf{u}(t)} \hat{\rho}\hat{A}^{(\alpha)\dagger}_{k,\mathbf{u}(t)} -\frac{1}{2}\hat{A}^{(\alpha)\dagger}_{k,\mathbf{u}(t)} \hat{A}^{(\alpha)}_{k,\mathbf{u}(t)} \hat{\rho} -\frac{1}{2} \hat{\rho}\hat{A}^{(\alpha)\dagger}_{k,\mathbf{u}(t)} \hat{A}^{(\alpha)}_{k,\mathbf{u}(t)}),
\end{multline}
where $\lambda_\alpha[d(t)] \in \{0,1\}$ are functions that determine which bath is coupled the QS, $\hat{A}^{(\alpha)}_{k,\mathbf{u}(t)}$ are the Lindblad operators, and $\gamma^{(\alpha)}_{k,\mathbf{u}(t)}$ are the corresponding rates.
In particular, $\lambda_\mathrm{H}(\mathrm{Hot}) = 1$, $\lambda_\mathrm{C}(\mathrm{Hot})=0$, while $\lambda_\mathrm{H}(\mathrm{Cold}) = 0$, $\lambda_\mathrm{C}(\mathrm{Cold})=1$, and $\lambda_\mathrm{H}(\mathrm{None})=\lambda_\mathrm{C}(\mathrm{None}) = 0$.
Notice that both the Lindblad operators and the rates can depend on time through the value of the control $\mathbf{u}(t)$. Their explicit form depends on the details of the system, i.e. on the Hamiltonian describing the dynamics of the overall system including the bath and the system-bath interaction. Below, we provide the explicit form of $\hat{A}^{(\alpha)}_{k,\mathbf{u}(t)}$ and $\gamma^{(\alpha)}_{k,\mathbf{u}(t)}$ used to model the three setups considered in the manuscript. 
We adopt the standard approach to compute the instantaneous power and heat currents \cite{alicki1979}
\begin{equation}
\begin{aligned}
	P(t) &\equiv -\Tr\left[\hat{\rho}(t)\, \frac{\partial}{\partial t}\hat{H}[\mathbf{u}(t)]\right] , \\
	J_\alpha(t) &\equiv \Tr\left[ \mathcal{D}^{(\alpha)}_{\mathbf{u}(t),d(t)}[\hat{\rho}(t)]\,{\hat{H}}[\mathbf{u}(t)]\right].
\end{aligned}
\end{equation}

In the two-level system heat engine, we consider the following Lindblad operators and corresponding rates (identifying $k= \pm$):
\begin{align}
	\hat{A}^{(\alpha)}_{\pm,u(t)} &= \hat{\sigma}_\pm, & \gamma^{(\alpha)}_{\pm,u(t)}  &= \Gamma_\alpha\, f(\pm\beta_\alpha u(t) E_0 ),
\end{align}
where $\hat{\sigma}_+$ and $\hat{\sigma}_-$ denote the raising and lowering operators, $\Gamma_\alpha$ is a constant rate which sets the thermalization timescale when the QS is coupled to bath $\alpha$, and $f(x) = [1+\exp(x)]^{-1}$ is the Fermi distribution. 
This choice can be derived, for example, when considering the qubit as a single-level quantum dot tunnel-coupled to metallic leads, with flat density of states, which act as heat baths \cite{beenakker1991,esposito2009,nazarov2009,erdman2017}. 

In the superconducting qubit refrigerator, we employ the model first put forward in Ref.~\cite{karimi2016}, and further studied in Refs.~\cite{pekola2019,funo2019}. In particular, we consider the following Lindblad operators and corresponding rates (identifying $k= \pm$):
\begin{equation}
\begin{aligned}
	\hat{A}^{(\alpha)}_{+,u(t)} &= -i\ket*{e_{u(t)}}\bra*{g_{u(t)}}, \\
	\hat{A}^{(\alpha)}_{-,u(t)} &= +i\ket*{g_{u(t)}}\bra*{e_{u(t)}},
\end{aligned}
\end{equation}
where $|g_{u(t)}\rangle$ and $|e_{u(t)}\rangle$ are, respectively, the instantaneous ground state and excited state of Eq.~(\ref{eq:h_fridge}). The corresponding rates are given by $\gamma^{(\alpha)}_{\pm,u(t)} = S_{\alpha}[\pm\Delta \epsilon_{u(t)}] $, where $\Delta \epsilon_{u(t)}$ is the instantaneous energy gap of the system, and
\begin{equation}
	S_\alpha(\Delta \epsilon)= \frac{g_{\alpha}}{2} \frac{1}{1+Q_\alpha^2( \Delta\epsilon/\omega_\alpha - \omega_\alpha/\Delta \epsilon )^2 } \frac{\Delta \epsilon}{e^{\beta_\alpha\Delta\epsilon}-1}
\end{equation}
is the noise power spectrum of bath $\alpha$. Here $\omega_\alpha$, $Q_\alpha$ and $g_\alpha$ are the base resonance frequency, quality factor and coupling strength of the resonant circuit acting as bath $\alpha=\mathrm{H},\mathrm{C}$ (see Refs.~\cite{karimi2016,funo2019} for details). As in Ref.~\cite{funo2019}, we choose $\omega_\mathrm{C}=2E_0\Delta$ and $\omega_\mathrm{H}=2E_0\sqrt{\Delta^2 +1/4}$, such that the C (H) bath is in resonance with the qubit when $u=0$ ($u=1/2$). The width of the resonance is governed by $Q_\alpha$. 
The total coupling strength to bath $\alpha$, plotted in Fig.~\ref{fig:qubit_fridge}d, is quantified by 
\begin{equation}
	\gamma^{(\alpha)}_{u(t)} \equiv \gamma^{(\alpha)}_{+,u(t)} + \gamma^{(\alpha)}_{-,u(t)}.
\end{equation}

In the Harmonic oscillator heat engine, following Ref.~\cite{rezek2006}, we describe the coupling to the baths through the Lindblad operators  $\hat{A}^{(\alpha)}_{+,u(t)} = \hat{a}_{u(t)}^\dagger$, $\hat{A}^{(\alpha)}_{-,u(t)} = \hat{a}_{u(t)}$ and corresponding rates $\gamma^{(\alpha)}_{+,u(t)} = \Gamma_\alpha \,n(\beta_\alpha u(t)\omega_0)$ and $\gamma^{(\alpha)}_{-,u(t)} = \Gamma_\alpha[1+ n(\beta_\alpha u(t) \omega_0 )]$, where we identify $k= \pm$. $\hat{a}_{u(t)}=(1/\sqrt{2})\sqrt{m\omega_0 u(t)}\,\hat{q} + i/\sqrt{m\omega_0 u(t)}\,\hat{p}$ and $\hat{a}_{u(t)}^\dagger$ are respectively the (control dependent) lowering and raising operators, $\Gamma_{\alpha}$ is a constant rate setting the thermalization timescale of the system coupled to bath $\alpha$, and $n(x)=[\exp(x)-1]^{-1}$ is the Bose-Einstein distribution.

\paragraph*{Reinforcement-Learning Algorithm.}
As discussed in the Results section, the choice of the reward as in Eq.~(\ref{eq:r_qtm}) guarantees that the aim of the RL agent is to maximize $\langle P_{[\nu ]} \rangle $ introduced in Eq.~(\ref
  {eq:avg_p_def}). To be precise, plugging Eq.~(\ref {eq:r_qtm}) into Eq.~(\ref
  {eq:return_def}) gives $\langle P_{[\nu ]} \rangle $  (up to an irrelevant constant prefactor) only in the limit of
  $\Delta t \to 0$. However, also for finite $\Delta t$, both quantities are
  time-averages of the power, so they are equally valid definitions to describe
  a long-term power maximization.
  
We use a generalization of the soft-actor critic (SAC) method, first developed for continuous actions
\cite{haarnoja2018_pmlr,haarnoja2018_arxiv_sac}, to handle a combination of discrete and continuous actions \cite{christodoulou2019,delalleau2019}. 
We here present an overview of our implementation of SAC putting special emphasis on the differences with respect to the standard implementation. However, we refer to \cite{haarnoja2018_pmlr,haarnoja2018_arxiv_sac,christodoulou2019,delalleau2019} for additional details. Our method, implemented with PyTorch, is based on modifications and generalizations of the SAC implementation provided by Spinning Up from OpenAI \cite{spinningup2018}. All code and data to reproduce the experiments is available online (see Data Availability and Code Availability sections).

The SAC algorithm is based on policy iteration, i.e. it consists of iterating multiple times over two steps: a \textit{policy evaluation step}, and a \textit{policy improvement step}. In the policy evaluation step, the value function of the current policy is (partially) learned, whereas in the policy improvement step a better policy is learned by making use of the value function. We now describe these steps more in detail.

In typical RL problems, the optimal policy $\pi^*(s|a)$ is defined as the policy that maximizes the expected reward defined in Eq.~(\ref{eq:return_def}), i.e.:
\begin{equation}
    \pi^* = \mathrm{arg}\max_\pi\, \mathop{\mathrm{E}_\pi}\limits_{ s\sim \mu_\pi}  \Big[ \sum_{k=0}^{\infty} \gamma^k \,r_{k+1} \Big| s_0 = s \Big],
    \label{eq:pi_star_rl}
\end{equation}
where $\mathrm{E}_\pi$ denotes the expectation value choosing actions according to the policy $\pi$. The initial state $s_0=s$ is sampled from $\mu_\pi$, i.e. the steady-state distribution of states that are visited by $\pi$. 
In the SAC method, balance between exploration and exploitation \cite{sutton2018} is achieved by introducing an Entropy-Regularized maximization objective. In this setting, the optimal policy $\pi^*$ is given by
\begin{equation}
    \pi^* = \mathrm{arg}\max_\pi\, \mathop{\mathrm{E}_\pi}\limits_{s\sim \mathcal{B}}\Big[ \sum_{k=0}^{\infty} \gamma^k \,\Big(r_{k+1} + \varepsilon H(\pi(\cdot|s_k))  \Big) \Big| s_0 =s  \Big],
    \label{eq:pi_star}
\end{equation}
where $\varepsilon \geq 0$ (usually denoted with $\alpha$ in the RL literature) is a hyper-parameter that balances the trade-off between exploration and exploitation, and
\begin{equation}
    H(P) = \mathop{\mathrm{E}}\limits_{x\sim P}[ -\log P(x) ]
\end{equation}
is the entropy of the probability distribution $P$. Notice that we replaced the unknown state distribution $\mu_\pi$ with $\mathcal{B}$, which is a replay buffer populated during training by storing the observed one-step transitions $(s_k,a_k, r_{k+1}, s_{k+1})$.

We define the value function $Q^\pi(s,a)$ of a given policy $\pi$ as
\begin{multline}
    Q^\pi(s,a) = \\
    \text{E}_\pi  \Big[ r_{1} + 
    \sum_{k=1}^{\infty} \gamma^k \,\Big(r_{k+1} + \varepsilon H(\pi(\cdot|s_k))  \Big) \Big| s_0=s, a_0=a \Big],
\end{multline}
and its recursive Bellman equation reads
\begin{multline}
    Q^\pi(s,a) =  \underset{ a_1 \sim \pi(\cdot|s_1) }{\text{E}} \Big[ r_{1} + \\
    \gamma \Big( Q^\pi(s_1,a_1) - \varepsilon \log\pi(a_1|s_1)  \Big) \Big| s_0=s, a_0=a \Big].
    \label{eq:bellman}
\end{multline}
We use a function approximator  $Q_\phi(s,a)$ (e.g. a neural network) to describe the value function of the current policy, where $\phi$ represents a collection of learnable parameters.

Next, we describe the policy $\pi(a|s)$. Since we are dealing with a combination of a discrete and continuous actions, we define $a = (u,d)$, where $u$ is the continuous action and $d$ is the discrete action (for simplicity, we describe the case of a single continuous action, though the generalization to multiple variables is straightforward). From now on, all functions of $a$ are also to be considered as functions of $u,d$. We decompose the joint probability distribution as
\begin{equation}
    \pi_\theta(u,d|s) = \pi_{\theta_\mathrm{D}}(d|s) \cdot \pi_{\theta_{\mathrm{U},d}}(u|d,s),
    \label{eq:pi_decomp}
\end{equation}
where $\pi_{\theta_\mathrm{D}}(d|s)$ is a function approximator for the marginal probability of taking discrete action $d$, which depends on learnable parameters $\theta_\mathrm{D}$, and $\pi_{\theta_{\mathrm{U},d}}(u|d,s)$ is a parameterization of the conditional probability density of choosing action $u$, given action $d$, which depends on learnable parameters $\theta_{\mathrm{U},d}$ - one set for each discrete action $d$. We denote with $\theta$ the collection of all parameters $\theta_\mathrm{D}$ and $\theta_{\mathrm{U},d}$. Notice that this decomposition allows us to describe correlations between the discrete and the continuous action, which are crucial in our application. We further parameterize $\pi_{\theta_{\mathrm{U},d}}(u|d,s)$ as a squashed Gaussian policy, i.e. as the distribution of the variable
\begin{equation}
\begin{aligned}
&
\begin{multlined}
 \tilde{u}_{\theta_{\text{U},d}}(s,\xi) = \\
    u_\text{a} + \frac{u_\text{b} - u_\text{a}}{2}[1+ \tanh\left( \mu_{\theta_{\text{U},d}}(s) + \sigma_{\theta_{\text{U},d}}(s)\cdot \xi )  \right)],   
\end{multlined}
     \\ & \xi \sim \mathcal{N}(0,1),
\end{aligned}
\label{eq:u_tilda}
\end{equation}
where $\mu_{\theta_{\mathrm{U},d}}(s)$ and $\sigma_{\theta_{\mathrm{U},d}}(s)$, representing respectively the mean and standard deviation of the Gaussian distribution, are function approximators which depend on the learnable parameters $\theta_{\mathrm{U},d}$, $\mathcal{N}(0,1)$ is the normal distribution with zero mean and unit variance, and where we assume that $\mathcal{U}=[u_\mathrm{a},u_\mathrm{b}]$. 

We now describe the policy evaluation step. In the SAC algorithm, we learn two value functions $Q_{\phi_i}(s,a)$ described by the learnable parameters $\phi_i$, for $i=1,2$. Since $Q_{\phi_i}(s,a)$ should satisfy the Bellman Eq.~(\ref{eq:bellman}), we define the loss function for $Q_{\phi_i}(s,a)$ as the mean square difference between the left and right hand side of Eq.~(\ref{eq:bellman}), i.e.
\begin{equation}
    L_Q(\phi_i) = \mathop{\mathrm{E}}\limits_{(s,a,r,s^\prime)\sim \mathcal{B}} \left[ ( Q_{\phi_i}(s,a) - y(r,s^\prime))^2  \right],
    \label{eq:q_loss}
\end{equation}
where 
\begin{multline}
    y(r,s^\prime) = r + \\ \gamma \underset{a^\prime \sim \pi_\theta(\cdot|s^\prime)}{\text{E}} \Big[ \min_{j=1,2}Q_{\phi_{\text{targ},j}}(s^\prime,a^\prime) - \varepsilon \log \pi_\theta(a^\prime|s^\prime) \Big].
    \label{eq:y_1}
\end{multline}
Notice that in Eq.~(\ref{eq:y_1}) we replaced $Q_{\phi_i}$ with $\min_{j=1,2}Q_{\phi_{\mathrm{targ},j}}$, where $\phi_{\mathrm{targ},j}$, for $j=1,2$, are target parameters which are not updated when minimizing the loss function; instead, they are held fixed during backpropagation, and then they are updated according to polyak averaging, i.e.
\begin{equation}
    \phi_{\mathrm{targ},i} \leftarrow \rho_\mathrm{polyak} \phi_{\mathrm{targ},i} + (1-\rho_\mathrm{polyak})\phi_{i},
\end{equation}
where $\rho_\mathrm{polyak}$ is a hyperparameter. This change was shown to improve learning \cite{haarnoja2018_pmlr,haarnoja2018_arxiv_sac}. In order to evaluate the expectation value in Eq.~(\ref{eq:y_1}), we use the decomposition in Eq.~(\ref{eq:pi_decomp}) to write
\begin{equation}
    \mathop{\mathrm{E}}\limits_{ a^\prime\sim \pi_\theta(\cdot|s^\prime)}[\cdot] =
   \sum_{{d}^\prime} \pi_{\theta_\mathrm{D}}({d}^\prime|s^\prime) \mathop{\mathrm{E}}\limits_{{u}^\prime \sim \pi_{\theta_{U,{d}^\prime}}(\cdot|d^\prime,s^\prime) }[\cdot] ,
   \label{eq:e_decomp}
\end{equation}
where we denote $a^\prime = (u^\prime, d^\prime)$. Plugging Eq.~(\ref{eq:e_decomp}) into Eq.~(\ref{eq:y_1}), and approximating the expectation value over $u^\prime$ with a single sampled value  yields 
\begin{equation}
\begin{aligned}
&
\begin{multlined}
    y(r,s^\prime) = r + \gamma \sum_{{d}^\prime} \pi_{\theta_\text{D}}({d}^\prime|s^\prime) \times \\
    \Big[ \min_{j=1,2}Q^\pi_{\phi_{\text{targ},j}}(s^\prime,d^\prime,u^\prime) - \alpha \log \pi_\theta(d^\prime,u^\prime|s^\prime) \Big],
    \label{eq:y_2}
\end{multlined}
\\ &u^\prime \sim \pi_{\theta_{\text{U},d^\prime}}(\cdot|{d}^\prime,s^\prime).
\end{aligned}
\end{equation}
We therefore perform a full average over the discrete action, and a single sampling of the continuous action.

We now turn to the policy improvement step. Given a policy $\pi_{\theta_{\mathrm{old}}}$, Ref.~\cite{haarnoja2018_arxiv_sac} proves that $\pi_{\theta_{\mathrm{new}}}$ is a better policy [respect to maximization in Eq.~(\ref{eq:pi_star})] if we update the policy parameters according to
\begin{equation}
    \theta_\mathrm{new} = \mathrm{arg}\min_\theta D_\mathrm{KL}\Big( \pi_\theta(\cdot|s)   \Big|\Big| \frac{\exp\left( Q^{\pi_{\theta_\mathrm{old}}}(s,\cdot)/\alpha \right) }{ Z^{\pi_{\theta_\mathrm{old}}}}  \Big),
    \label{eq:theta_new}
\end{equation}
where $s$ is any state, $D_\mathrm{KL}$ denotes the Kullback-Leibler divergence, and $Z^{\pi_{\theta_\mathrm{old}}}$ is the partition function of the exponential of the value function. Intuitively, this step is the equivalent of making the policy $\epsilon$-greedy in the standard RL setting. The idea is to use the minimization in Eq.~(\ref{eq:theta_new}) to define a loss function to perform an update of $\theta$. Noting that the partition function does not impact the gradient, multiplying the Kullback-Leibler divergence by $\alpha$, and replacing $Q^{\pi_{\theta_\mathrm{old}}}$ with $\min_j Q_{\phi_j}$, we define the loss function as
\begin{equation}
    L_\pi(\theta) = \mathop{\mathrm{E}}\limits_{{s\sim \mathcal{B} \atop   a\sim \pi_\theta(\cdot|s)}}\left[ \alpha \log\pi_\theta(a|s) - \min_{j=1,2} Q_{\phi_j}(s,a) \right].
    \label{eq:l_1}
\end{equation}
In order to evaluate the expectation value in Eq.~(\ref{eq:l_1}), we use the previous trick of averaging the discrete action, and performing a single sample of the continuous action using $\xi$. Recalling Eq.~(\ref{eq:u_tilda}), this yields
\begin{equation}
\begin{aligned}
&
\begin{multlined}
    L_\pi(\theta) = \underset{ \substack{s\sim \mathcal{B}  }}{\text{E}}\Big[\sum_d \pi_{\theta_\text{D}}(d|s) \Big(  \alpha \log\pi_\theta(d,\tilde{u}_{\theta_{\text{U},d}}(s,\xi)|s) - \\
    \min_{j=1,2} Q_{\phi_j}(s,d,\tilde{u}_{\theta_{\text{U},d}}(s,\xi))  \Big) \Big],
    \end{multlined}
    \\ &\xi \sim \mathcal{N}(0,1).
\end{aligned}
    \label{eq:pi_loss}
\end{equation}

To summarize, the SAC algorithm consists of repeating over and over a policy evaluation step, and a  policy improvement step. The policy evaluation step consists of a single optimization step to minimize the loss functions $L_Q(\phi_i)$ (for $i=1,2$), given in Eq.~(\ref{eq:q_loss}), where $y(r,s^\prime)$ is computed using Eq.~(\ref{eq:y_2}).
The policy improvement step consists of a single optimization step to minimize the loss function $L_\pi(\theta)$ given in Eq.~(\ref{eq:pi_loss}). In both loss functions, the expectation value over the states is approximated with a batch of experience sampled randomly from the replay buffer $\mathcal{B}$.

\paragraph*{Training details.}
We now provide the details of the algorithm used to learn the four specific cycles described in the main manuscript. 

The value function $Q_\phi(s,u,d)$ is parameterized the following way. We use a fully connected neural network (NN), with two hidden layers, that takes $s$ and $u$ as input (by stacking them into a single array), and outputs $|D|$ values, where $|D|$ is the number of discrete actions. The $i_{th}$ output corresponds to $Q_\phi(s,u,d=d_i)$, where $\{d_i\}$ are the possible discrete actions. We use the ReLU activation function in all layers except for the output layer, were we apply the identity (since the value function can take arbitrary positive or negative values). The parameters $\phi$ correspond to the weights and biases of the whole network.

The policy $\pi_\theta(u,d|s) = \pi_{\theta_\mathrm{D}}(d|s)\cdot \pi_{\theta_{\mathrm{U},d}}(u,|d,s)$ is parameterized the following way. We use a fully connected NN, with two hidden layers, that takes $s$ as input, and produces $3\cdot |D|$ values, corresponding to
\begin{equation}
    \left\{ \pi_{\theta_\mathrm{D}}(d_i|s), \,\, \mu_{\theta_{\mathrm{U},d_i}}(s), \,\, \sigma_{\theta_{\mathrm{U},d_i}}(s)) \right\}_{i=1,\dots,|D|}.
\end{equation} 
More specifically, we use the ReLU activation in all layers except for the output layer, where we use the identity. However, in order to enforce the normalization $\sum_i \pi_{\theta_\mathrm{D}}(d_i|s) =1$, we apply a soft-max to the corresponding outputs, and instead of outputting $\sigma_{\theta_{\mathrm{U},d_i}}(s)$, we output $\log(\sigma_{\theta_{\mathrm{U},d_i}}(s))$, which has no constraint on the sign.

In order to enforce sufficient exploration in the early stage of training, we do the following. For a fixed number of initial steps, we choose random actions sampling them uniformly withing their range. Furthermore, for another fixed number of initial steps, we do not update the parameters to allow the replay buffer to have enough transitions. $\mathcal{B}$ is a first-in-first-out buffer, of fixed dimension, from which batches of transitions are randomly sampled to update the NN parameters. After this initial phase, we repeat a policy evaluation and a policy improvement step $n_\mathrm{updates}$ times every $n_\mathrm{updates}$ steps. This way, the overall number of updates coincides with the number of actions performed on the environment. The optimization steps are performed using the ADAM optimizer with the standard values of $\beta_1$ and $\beta_2$. To favor an exploratory behaviour early in the training, and at the same time to end up with a policy that is approximately deterministic, we schedule $\varepsilon$. In particular, we vary it during each step according to
\begin{equation}
    \varepsilon(n_\mathrm{steps}) = \varepsilon_0\, \exp(-n_\mathrm{steps}/\varepsilon_\mathrm{decay}),
\end{equation}
where $n_\mathrm{steps}$ is the current step number, and $\varepsilon_0$ and $\varepsilon_\mathrm{decay}$ are hyperparameters.

All hyperparameters used to produce the cycles in Figs.~\ref{fig:qubit_engine}, \ref{fig:qubit_fridge} and \ref{fig:harmonic_engine} are provided in Table~\ref{tab:hyper}.
\begin{table}[h]
\centering
\begin{tabular}{lccc}
\toprule
Hyperparameter & ~ Figs.~\ref{fig:qubit_engine}, \ref{fig:qubit_fridge} ~ & Fig.~\ref{fig:harmonic_engine}a ~ & Fig.~\ref{fig:harmonic_engine}b  \\
\midrule
Hidden layers  &2 &2 &2\\
Hidden layer units & 256& 256& 256\\
Initial random steps  &5k &5k &10k\\
First update at step  &1k &1k &1k\\
Batch size  &256 &256 &256 \\
learning rate & 0.001& 0.001& 0.0005 \\
$\varepsilon_0$  &50 &50 &300 \\
$\varepsilon_\text{decay}$  &48k &24k &48k \\
$n_\text{updates}$  &50 &50 &50  \\
$\rho_\text{polyak}$   &0.995 &0.995 &0.995 \\
$\mathcal{B}$ size  &192k &192k &192k \\
\bottomrule
\end{tabular}
\caption{\textbf{Hyperparameters used in numerical calculations that are not reported in the Figure captions.}}
\label{tab:hyper}
\end{table}

At last, we discuss the parameterization of the state $s$. As discussed in the manuscript, we use $s = (\hat{\rho},u)$ as state. While $u$ is passed to the NNs as-is, we now detail how we encoded $\hat{\rho}$ in the 3 systems studied in the manuscript.

In the two-level system heat engine, a closed equation of motion governing the evolution of $p(t) \equiv \mathrm{Tr}[\hat{\rho}(t) \hat{\sigma}_+\hat{\sigma}_- ]$ can be derived from the Markovian master equation, and the instantaneous heat flux $J_\alpha(t)$ can be expressed solely in terms of $p(t)$ (see Ref.~\cite{erdman2019_njp} for details). Therefore, we use the single parameter $p(t) \in [0,1]$ to encode $\hat{\rho}$.

In the superconducting qubit refrigerator, we encode $\hat{\rho}$ using the following three real parameters: $\langle  e_{u(t)} | \hat{\rho}(t) | e_{u(t)} \rangle$, $\Re[\langle g_{u(t)}|\hat{\rho}(t)|e_{u(t)}\rangle]$ and $\Im[\langle g_{u(t)}|\hat{\rho}(t)|e_{u(t)}\rangle]$. These fully characterize a density matrix of a qubit.

In the Harmonic oscillator heat engine a closed set of equations of motion can be derived from the Markovian master equation for the following three quantities:
\begin{eqnarray}
    & & H(t) \equiv \mathrm{Tr}[\hat{\rho}(t) \hat{H}[u(t)]], \nonumber \\
    & & L(t) \equiv \mathrm{Tr}[\hat{\rho}(t) (  \frac{1}{2m} \hat{p}^2 - \frac{1}{2}m (u(t)w_0)^2 \hat{q}^2   )  ], \nonumber \\
    & & D(t) \equiv \mathrm{Tr}[\hat{\rho}(t) ( \hat{q}\hat{p}+\hat{p}\hat{q} ) ].
\end{eqnarray}
Furthermore, the heat flux $J_\alpha(t)$ can be expressed solely in terms of these quantities (see Ref. \cite{rezek2006} for details). We thus use these three quantities to encode $\hat{\rho}$. More specifically, since they are not bounded in an obvious way by some system parameter, for numerical stability we use $\tilde{O}(t) \equiv \log(|O(t)| + \delta)$ and $sO(t) = \mathrm{sign}(O(t))$ instead of $O= H, L, D$ to encode the state, where $\delta=10^{-20}$ is a small parameter introduced to prevent numerical divergences. At last, we do not use $sH(t)$ since $H(t)$ is always a positive quantity (being the energy of the harmonic oscillator). Therefore, we encode $\hat{\rho}$ using the following 5 parameters: $(\tilde{H},\tilde{L},\tilde{D},sL,sD)$.

\paragraph*{Convergence of the RL approach.} The training process presents some degree of stochasticity, such as the initial random steps, and the random sampling of a batch of experience from the replay buffer to compute an approximate gradient of the loss functions. We thus need to evaluate the reliability of our approach.

\begin{figure}[!tb]
	\centering
	\includegraphics[width=0.99\columnwidth]{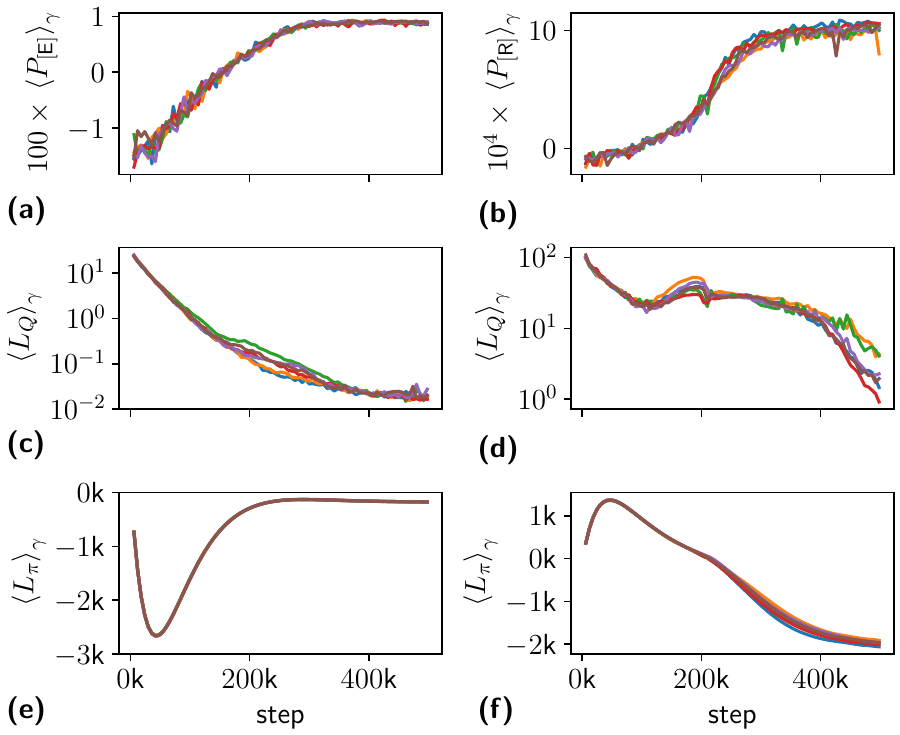}
	\caption{\textbf{Training curves}. Training curves for $6$ consecutive runs of the RL method in the two-level system heat engine case [panels (a), (c) and (e), corresponding to Fig.~\ref{fig:qubit_engine}] and in the superconducting qubit refrigerator case [panels (b), (d) and (f), corresponding to Fig.~\ref{fig:qubit_fridge}]. The panels show, as a function of the training steps, the running average of the power [panels (a) and (b)], the running average of the loss function $L_Q$ [panels (c) and (d)], and the running average of the loss function $L_\pi$ [panels (e) and (f)]. Each curve corresponds to a separate training.}
	\label{fig:two_level_qubit_convergence}
\end{figure}
\begin{figure}[!tb]
	\centering
	\includegraphics[width=0.99\columnwidth]{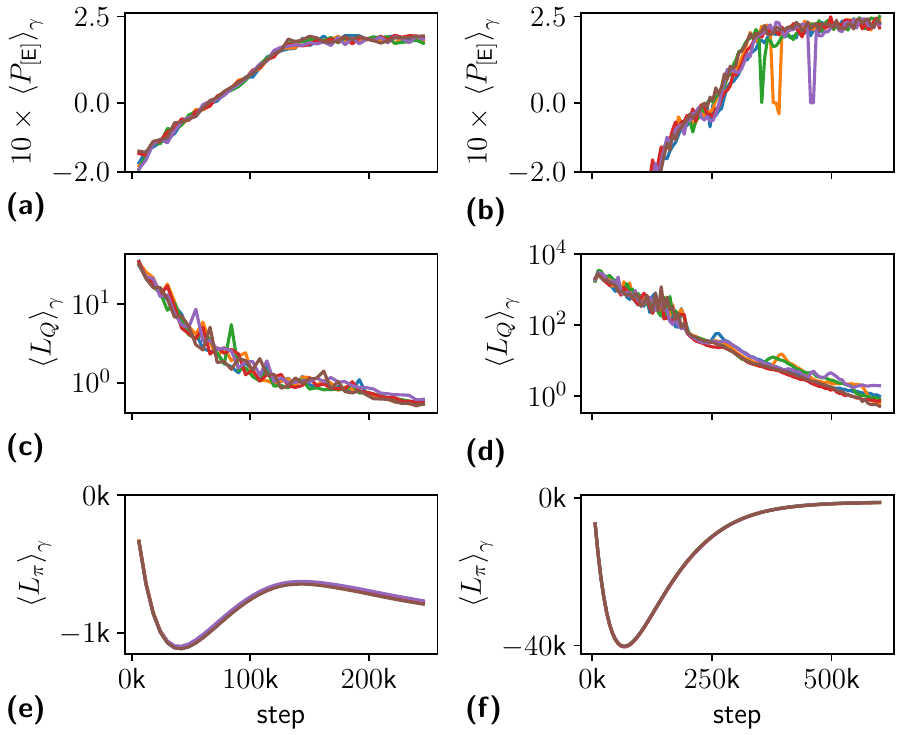}
	\caption{\textbf{Training curves}. Training curves for $6$ consecutive runs of the RL method in the harmonic oscillator heat engine. Panels (a), (c) and (e) correspond to the optimization carried out in the range considered in Figs.~\ref{fig:harmonic_engine}a and \ref{fig:harmonic_engine}c, while panels (b), (d) and (f)  correspond to the optimization carried out in Figs.~\ref{fig:harmonic_engine}b and \ref{fig:harmonic_engine}d. The panels show, as a function of the training steps, the running average of the power [panels (a) and (b)], the running average of the loss function $L_Q$ [panels (c) and (d)], and the running average of the loss function $L_\pi$ [panels (e) and (f)]. Each curve corresponds to a separate training.}
	\label{fig:harmonic_convergence}
\end{figure}
In Figs.~\ref{fig:two_level_qubit_convergence} an \ref{fig:harmonic_convergence}, we show the training curves for $6$ consecutive runs of our method applied to all $4$ cases studied in the main text. 
More specifically, panels (a), (c) and (e) of Fig.~\ref{fig:two_level_qubit_convergence} correspond to the training of the two-level system heat engine (considered in Fig.~\ref{fig:qubit_engine}), while panels (b), (d) and (f) of Fig.~\ref{fig:two_level_qubit_convergence} correspond to the training of the superconducting qubit refrigerator (considered in Fig.~\ref{fig:qubit_fridge}). Panels (a), (c) and (e) of Fig.~\ref{fig:harmonic_convergence} correspond to the training of the Harmonic oscillator in the interval considered in Figs.~\ref{fig:harmonic_engine}a and \ref{fig:harmonic_engine}c, while panels (b), (d) and (f) of Fig.~\ref{fig:harmonic_convergence} correspond to the training of the Harmonic oscillator in the interval considered in Figs.~\ref{fig:harmonic_engine}b and \ref{fig:harmonic_engine}d. 
Both Figs.~\ref{fig:two_level_qubit_convergence} and \ref{fig:harmonic_convergence} show, as a function of the step, the running average of the power [panels (a) and (b)], the running average $\ev{L_Q}_\gamma$ of the loss function $L_Q$ [panels (c) and (d)] and the running average $\ev{L_\pi}_\gamma$ of the loss function $L_\pi$ [panels (e) and (f)] computed on the batch of experience used to estimate the gradient of the corresponding loss function. Every curve corresponds to a separate training. 

As we can see in Figs.~\ref{fig:two_level_qubit_convergence} an \ref{fig:harmonic_convergence}, in all $4$ cases the running average of the reward reliably converges to a solution yielding similar values of the power, and also the running averages of the loss functions display a qualitatively similar behavior. We notice that while $L_Q$ is the mean square of the Bellman error [see Eq.~(\ref{eq:q_loss})], $L_\pi$ is just a function whose gradient provides a better policy, so its value during training is not required to be a decreasing function.

\paragraph*{Generation of coherence.}
In order to quantify the coherence generated in the instantaneous eigenbasis of the Hamiltonian in the refrigerator based on a superconducting qubit, we evaluated the time average of \textit{relative entropy of coherence} \cite{baumgratz2014}, defined as
\begin{equation}
	C(\hat{\rho}(t)) = S(\hat{\rho}_\mathrm{diag.}(t)) -  S(\hat{\rho}(t)),
\end{equation}
where $S(\hat{\rho}) = -\mathrm{Tr}[\hat{\rho}\ln\hat{\rho}]$ is the Von Neumann entropy, and
\begin{multline}
	\hat{\rho}_\mathrm{diag.}(t) = \langle g_{u(t)} | \hat{\rho}(t)| g_{u(t)}\rangle \cdot | g_{u(t)}\rangle \langle g_{u(t)} | \\ +  \langle e_{u(t)} | \hat{\rho}(t)| e_{u(t)}\rangle \cdot | e_{u(t)}\rangle \langle e_{u(t)} |
\end{multline}
is the density matrix, in the instantaneous eigenbasis $|g_{u(t)}\rangle$ and $|e_{u(t)}\rangle$, with the off-diagonal terms canceled out. We find that that the time-average of the relative entropy of coherence is $\approx 0.116$ in the cycle found by the RL agent, while it is $\approx 0.194$ applying the trapezoidal cycle with the same period.

\paragraph*{Otto cycle comparison.}
In Fig.~\ref{fig:harmonic_engine} we compare the performance of an Otto cycle, optimized as in the upper panel of Fig. 6 of Ref.~\cite{rezek2006}, with the cycle discovered by the RL agent. However, Ref.~\cite{rezek2006} only provides the value of the power of the optimized Otto cycle, not the durations of the four strokes that produce such power. We therefore performed a grid search in the space of these four durations. After identifying the largest power, we ran the Netwon algorithm to further maximize the power. The final cycle we found is the one shown as a dashed lines in Figs.~\ref{fig:harmonic_engine}c and \ref{fig:harmonic_engine}d. The corresponding power, shown as a dashed line in Figs.~\ref{fig:harmonic_engine}a and \ref{fig:harmonic_engine}b, nicely matches with Ref.~\cite{rezek2006}.

\section*{Data Availability}
All raw data was generated with the accompanying code and is available in Figshare (\url{https://doi.org/10.6084/m9.figshare.16822054.v1}).

\section*{Code Availability}
The code used to generate all results is available on GitHub (\url{https://github.com/PaoloAE/paper_rl_thermal_machines}).

\section*{Acknowledgements} We gratefully acknowledge funding by the BMBF (Berlin Institute for the Foundations of Learning and Data -- BIFOLD), the European Research Commission (ERC CoG 772230) and the Berlin Mathematics Center MATH+ (AA1-6, AA2-8).

\section*{Competing interests}
P.A.E. and F.N. are authors of a patent application containing aspects of this work (Application to the European Patent Office, file number: 21 191 966.7).

\section*{Author Contributions}
P.A.E. and F.N. designed the research and method. P.A.E. wrote the computer code and carried out the numerical calculations. P.A.E. and F.N. analysed the data and wrote the manuscript.

\clearpage

%\bibliography{references}

\begin{thebibliography}{107}%
\makeatletter
\providecommand \@ifxundefined [1]{%
 \@ifx{#1\undefined}
}%
\providecommand \@ifnum [1]{%
 \ifnum #1\expandafter \@firstoftwo
 \else \expandafter \@secondoftwo
 \fi
}%
\providecommand \@ifx [1]{%
 \ifx #1\expandafter \@firstoftwo
 \else \expandafter \@secondoftwo
 \fi
}%
\providecommand \natexlab [1]{#1}%
\providecommand \enquote  [1]{``#1''}%
\providecommand \bibnamefont  [1]{#1}%
\providecommand \bibfnamefont [1]{#1}%
\providecommand \citenamefont [1]{#1}%
\providecommand \href@noop [0]{\@secondoftwo}%
\providecommand \href [0]{\begingroup \@sanitize@url \@href}%
\providecommand \@href[1]{\@@startlink{#1}\@@href}%
\providecommand \@@href[1]{\endgroup#1\@@endlink}%
\providecommand \@sanitize@url [0]{\catcode `\\12\catcode `\$12\catcode
  `\&12\catcode `\#12\catcode `\^12\catcode `\_12\catcode `\%12\relax}%
\providecommand \@@startlink[1]{}%
\providecommand \@@endlink[0]{}%
\providecommand \url  [0]{\begingroup\@sanitize@url \@url }%
\providecommand \@url [1]{\endgroup\@href {#1}{\urlprefix }}%
\providecommand \urlprefix  [0]{URL }%
\providecommand \Eprint [0]{\href }%
\providecommand \doibase [0]{https://doi.org/}%
\providecommand \selectlanguage [0]{\@gobble}%
\providecommand \bibinfo  [0]{\@secondoftwo}%
\providecommand \bibfield  [0]{\@secondoftwo}%
\providecommand \translation [1]{[#1]}%
\providecommand \BibitemOpen [0]{}%
\providecommand \bibitemStop [0]{}%
\providecommand \bibitemNoStop [0]{.\EOS\space}%
\providecommand \EOS [0]{\spacefactor3000\relax}%
\providecommand \BibitemShut  [1]{\csname bibitem#1\endcsname}%
\let\auto@bib@innerbib\@empty
%</preamble>
\bibitem [{\citenamefont {Fagas}\ \emph {et~al.}(2014)\citenamefont {Fagas},
  \citenamefont {Gammaitoni}, \citenamefont {Paul},\ and\ \citenamefont
  {Berini}}]{fagas2014}%
  \BibitemOpen
  \bibfield  {author} {\bibinfo {author} {\bibfnamefont {G.}~\bibnamefont
  {Fagas}}, \bibinfo {author} {\bibfnamefont {L.}~\bibnamefont {Gammaitoni}},
  \bibinfo {author} {\bibfnamefont {D.}~\bibnamefont {Paul}},\ and\ \bibinfo
  {author} {\bibfnamefont {G.~A.}\ \bibnamefont {Berini}},\ }\href@noop {}
  {\emph {\bibinfo {title} {ICT - Energy - Concepts Towards Zero Power
  Information and Communication Technology}}}\ (\bibinfo  {publisher}
  {InTech},\ \bibinfo {year} {2014})\BibitemShut {NoStop}%
\bibitem [{\citenamefont {Pekola}(2015)}]{pekola2015}%
  \BibitemOpen
  \bibfield  {author} {\bibinfo {author} {\bibfnamefont {J.~P.}\ \bibnamefont
  {Pekola}},\ }\bibfield  {title} {\bibinfo {title} {Towards quantum
  thermodynamics in electronic circuits},\ }\href
  {https://doi.org/10.1038/nphys3169} {\bibfield  {journal} {\bibinfo
  {journal} {Nat. Phys.}\ }\textbf {\bibinfo {volume} {11}},\ \bibinfo {pages}
  {118} (\bibinfo {year} {2015})}\BibitemShut {NoStop}%
\bibitem [{\citenamefont {Giazotto}\ \emph {et~al.}(2006)\citenamefont
  {Giazotto}, \citenamefont {Heikkil\"a}, \citenamefont {Luukanen},
  \citenamefont {Savin},\ and\ \citenamefont {Pekola}}]{giazotto2006}%
  \BibitemOpen
  \bibfield  {author} {\bibinfo {author} {\bibfnamefont {F.}~\bibnamefont
  {Giazotto}}, \bibinfo {author} {\bibfnamefont {T.~T.}\ \bibnamefont
  {Heikkil\"a}}, \bibinfo {author} {\bibfnamefont {A.}~\bibnamefont
  {Luukanen}}, \bibinfo {author} {\bibfnamefont {A.~M.}\ \bibnamefont
  {Savin}},\ and\ \bibinfo {author} {\bibfnamefont {J.~P.}\ \bibnamefont
  {Pekola}},\ }\bibfield  {title} {\bibinfo {title} {Opportunities for
  mesoscopics in thermometry and refrigeration: Physics and applications},\
  }\href {https://doi.org/10.1103/RevModPhys.78.217} {\bibfield  {journal}
  {\bibinfo  {journal} {Rev. Mod. Phys.}\ }\textbf {\bibinfo {volume} {78}},\
  \bibinfo {pages} {217} (\bibinfo {year} {2006})}\BibitemShut {NoStop}%
\bibitem [{\citenamefont {Binder}\ \emph {et~al.}(2019)\citenamefont {Binder},
  \citenamefont {Correa}, \citenamefont {Gogolin}, \citenamefont {Anders},\
  and\ \citenamefont {Adesso}}]{binder2019}%
  \BibitemOpen
  \bibfield  {author} {\bibinfo {author} {\bibfnamefont {F.}~\bibnamefont
  {Binder}}, \bibinfo {author} {\bibfnamefont {L.}~\bibnamefont {Correa}},
  \bibinfo {author} {\bibfnamefont {C.}~\bibnamefont {Gogolin}}, \bibinfo
  {author} {\bibfnamefont {J.}~\bibnamefont {Anders}},\ and\ \bibinfo {author}
  {\bibfnamefont {G.}~\bibnamefont {Adesso}},\ }\href@noop {} {\emph {\bibinfo
  {title} {Thermodynamics in the Quantum Regime: Fundamental Aspects and New
  Directions}}},\ Fundamental Theories of Physics\ (\bibinfo  {publisher}
  {Springer International Publishing},\ \bibinfo {year} {2019})\BibitemShut
  {NoStop}%
\bibitem [{\citenamefont {Vinjanampathy}\ and\ \citenamefont
  {Anders}(2016)}]{vinjanampathy2016}%
  \BibitemOpen
  \bibfield  {author} {\bibinfo {author} {\bibfnamefont {S.}~\bibnamefont
  {Vinjanampathy}}\ and\ \bibinfo {author} {\bibfnamefont {J.}~\bibnamefont
  {Anders}},\ }\bibfield  {title} {\bibinfo {title} {Quantum thermodynamics},\
  }\href {https://doi.org/10.1080/00107514.2016.1201896} {\bibfield  {journal}
  {\bibinfo  {journal} {Contemp. Phys.}\ }\textbf {\bibinfo {volume} {57}},\
  \bibinfo {pages} {545} (\bibinfo {year} {2016})}\BibitemShut {NoStop}%
\bibitem [{\citenamefont {Friedenauer}\ \emph {et~al.}(2008)\citenamefont
  {Friedenauer}, \citenamefont {Schmitz}, \citenamefont {Glueckert},
  \citenamefont {Porras},\ and\ \citenamefont {Schaetz}}]{friedenauer2008}%
  \BibitemOpen
  \bibfield  {author} {\bibinfo {author} {\bibfnamefont {H.}~\bibnamefont
  {Friedenauer}}, \bibinfo {author} {\bibfnamefont {H.}~\bibnamefont
  {Schmitz}}, \bibinfo {author} {\bibfnamefont {J.}~\bibnamefont {Glueckert}},
  \bibinfo {author} {\bibfnamefont {D.}~\bibnamefont {Porras}},\ and\ \bibinfo
  {author} {\bibfnamefont {T.}~\bibnamefont {Schaetz}},\ }\bibfield  {title}
  {\bibinfo {title} {Simulating a quantum magnet with trapped ions},\ }\href
  {https://doi.org/10.1038/nphys1032} {\bibfield  {journal} {\bibinfo
  {journal} {Nat. Phys.}\ }\textbf {\bibinfo {volume} {4}},\ \bibinfo {pages}
  {757} (\bibinfo {year} {2008})}\BibitemShut {NoStop}%
\bibitem [{\citenamefont {Blatt}\ and\ \citenamefont {Roos}(2012)}]{blatt2012}%
  \BibitemOpen
  \bibfield  {author} {\bibinfo {author} {\bibfnamefont {R.}~\bibnamefont
  {Blatt}}\ and\ \bibinfo {author} {\bibfnamefont {C.}~\bibnamefont {Roos}},\
  }\bibfield  {title} {\bibinfo {title} {Quantum simulations with trapped
  ions},\ }\href {https://doi.org/10.1038/nphys2252} {\bibfield  {journal}
  {\bibinfo  {journal} {Nat. Phys.}\ }\textbf {\bibinfo {volume} {8}},\
  \bibinfo {pages} {277} (\bibinfo {year} {2012})}\BibitemShut {NoStop}%
\bibitem [{\citenamefont {Childress}\ \emph {et~al.}(2006)\citenamefont
  {Childress}, \citenamefont {{Gurudev Dutt}}, \citenamefont {Taylor},
  \citenamefont {Zibrov}, \citenamefont {Jelezko}, \citenamefont {Wrachtrup},
  \citenamefont {Hemmer},\ and\ \citenamefont {Lukin}}]{childress2006}%
  \BibitemOpen
  \bibfield  {author} {\bibinfo {author} {\bibfnamefont {L.}~\bibnamefont
  {Childress}} \textit{et al.,} }\bibfield  {title} {\bibinfo
  {title} {Coherent dynamics of coupled electron and nuclear spin qubits in
  diamond},\ }\href {https://doi.org/10.1126/science.1131871} {\bibfield
  {journal} {\bibinfo  {journal} {Science}\ }\textbf {\bibinfo {volume}
  {314}},\ \bibinfo {pages} {281} (\bibinfo {year} {2006})}\BibitemShut
  {NoStop}%
\bibitem [{\citenamefont {Wallraff}\ \emph {et~al.}(2004)\citenamefont
  {Wallraff}, \citenamefont {Schuster}, \citenamefont {Blais}, \citenamefont
  {Frunzio}, \citenamefont {Huang}, \citenamefont {Majer}, \citenamefont
  {Kumar}, \citenamefont {Girvin},\ and\ \citenamefont
  {Schoelkopf}}]{wallraff2004}%
  \BibitemOpen
  \bibfield  {author} {\bibinfo {author} {\bibfnamefont {A.}~\bibnamefont
  {Wallraff}} \textit{et al.,} }\bibfield  {title} {\bibinfo {title} {Strong
  coupling of a single photon to a superconducting qubit using circuit quantum
  electrodynamics},\ }\href {https://doi.org/10.1038/nature02851} {\bibfield
  {journal} {\bibinfo  {journal} {Nature}\ }\textbf {\bibinfo {volume} {431}},\
  \bibinfo {pages} {162} (\bibinfo {year} {2004})}\BibitemShut {NoStop}%
\bibitem [{\citenamefont {Petta}\ \emph {et~al.}(2005)\citenamefont {Petta},
  \citenamefont {Johnson}, \citenamefont {Taylor}, \citenamefont {Laird},
  \citenamefont {Yacoby}, \citenamefont {Lukin}, \citenamefont {Marcus},
  \citenamefont {Hanson},\ and\ \citenamefont {Gossard}}]{peta2005}%
  \BibitemOpen
  \bibfield  {author} {\bibinfo {author} {\bibfnamefont {J.~R.}\ \bibnamefont
  {Petta}} \textit{et al.,} }\bibfield  {title} {\bibinfo {title} {Coherent
  manipulation of coupled electron spins in semiconductor quantum dots},\
  }\href {https://doi.org/10.1126/science.1116955} {\bibfield  {journal}
  {\bibinfo  {journal} {Science}\ }\textbf {\bibinfo {volume} {309}},\ \bibinfo
  {pages} {2180} (\bibinfo {year} {2005})}\BibitemShut {NoStop}%
\bibitem [{\citenamefont {Ronzani}\ \emph {et~al.}(2018)\citenamefont
  {Ronzani}, \citenamefont {Karimi}, \citenamefont {Senior}, \citenamefont
  {Chang}, \citenamefont {Peltonen}, \citenamefont {Chen},\ and\ \citenamefont
  {Pekola}}]{ronzani2018}%
  \BibitemOpen
  \bibfield  {author} {\bibinfo {author} {\bibfnamefont {A.}~\bibnamefont
  {Ronzani}} \textit{et al.,} }\bibfield  {title}
  {\bibinfo {title} {Tunable photonic heat transport in a quantum heat valve},\
  }\href {https://doi.org/10.1038/s41567-018-0199-4} {\bibfield  {journal}
  {\bibinfo  {journal} {Nat. Phys.}\ }\textbf {\bibinfo {volume} {14}},\
  \bibinfo {pages} {991} (\bibinfo {year} {2018})}\BibitemShut {NoStop}%
\bibitem [{\citenamefont {Dutta}\ \emph {et~al.}(2019)\citenamefont {Dutta},
  \citenamefont {Majidi}, \citenamefont {Corral}, \citenamefont {Erdman},
  \citenamefont {Florens}, \citenamefont {Costi}, \citenamefont {Courtois},\
  and\ \citenamefont {Winkelmann}}]{dutta2019}%
  \BibitemOpen
  \bibfield  {author} {\bibinfo {author} {\bibfnamefont {B.}~\bibnamefont
  {Dutta}} \textit{et al.,} }\bibfield  {title} {\bibinfo {title}
  {Direct probe of the seebeck coefficient in a kondo-correlated
  single-quantum-dot transistor},\ }\href
  {https://doi.org/10.1021/acs.nanolett.8b04398} {\bibfield  {journal}
  {\bibinfo  {journal} {Nano Lett.}\ }\textbf {\bibinfo {volume} {19}},\
  \bibinfo {pages} {506} (\bibinfo {year} {2019})}\BibitemShut {NoStop}%
\bibitem{maillet2020}%
  \BibitemOpen
  \bibfield  {author} {
  \bibinfo {author} {\bibfnamefont {O.}~\bibnamefont {Maillet}},
  \bibinfo {author} {\bibfnamefont {D.}~\bibnamefont {Subero}},
  \bibinfo {author} {\bibfnamefont {J.~T.}\ \bibnamefont {Peltonen}},
  \bibinfo {author} {\bibfnamefont {D.~S.}~\bibnamefont {Golubev}},\ and\ 
  \bibinfo {author} {\bibfnamefont {J.~P.}\ \bibnamefont {Pekola}},\ }\bibfield  {title}
  {\bibinfo {title} {Electric field control of radiative heat transfer in a superconducting circuit
},\ }\href {https://doi.org/10.1038/s41467-020-18163-8} {\bibfield
  {journal} {\bibinfo  {journal} {Nat. Commun.}\ }\textbf {\bibinfo {volume}
  {11}},\ \bibinfo {pages} {4326} (\bibinfo {year} {2020})}\BibitemShut {NoStop}%
\bibitem [{\citenamefont {Senior}\ \emph {et~al.}(2020)\citenamefont {Senior},
  \citenamefont {Gubaydullin}, \citenamefont {Karimi}, \citenamefont
  {Peltonen}, \citenamefont {Ankerhold},\ and\ \citenamefont
  {Pekola}}]{senior2020}%
  \BibitemOpen
  \bibfield  {author} {\bibinfo {author} {\bibfnamefont {J.}~\bibnamefont
  {Senior}} \textit{et al.,} }\bibfield  {title}
  {\bibinfo {title} {Heat rectification via a superconducting artificial
  atom},\ }\href {https://doi.org/10.1038/s42005-020-0307-5} {\bibfield
  {journal} {\bibinfo  {journal} {Commun. Phys.}\ }\textbf {\bibinfo {volume}
  {3}},\ \bibinfo {pages} {40} (\bibinfo {year} {2020})}\BibitemShut {NoStop}%
\bibitem [{\citenamefont {Ro{\ss}nagel}\ \emph {et~al.}(2016)\citenamefont
  {Ro{\ss}nagel}, \citenamefont {Dawkins}, \citenamefont {Tolazzi},
  \citenamefont {Abah}, \citenamefont {Lutz}, \citenamefont {Schmidt-Kaler},\
  and\ \citenamefont {Singer}}]{rossnagel2016}%
  \BibitemOpen
  \bibfield  {author} {\bibinfo {author} {\bibfnamefont {J.}~\bibnamefont
  {Ro{\ss}nagel}} \textit{et al.,} }\bibfield  {title} {\bibinfo
  {title} {A single-atom heat engine},\ }\href
  {https://doi.org/10.1126/science.aad6320} {\bibfield  {journal} {\bibinfo
  {journal} {Science}\ }\textbf {\bibinfo {volume} {352}},\ \bibinfo {pages}
  {325} (\bibinfo {year} {2016})}\BibitemShut {NoStop}%
\bibitem [{\citenamefont {Josefsson}\ \emph {et~al.}(2018)\citenamefont
  {Josefsson}, \citenamefont {Svilans}, \citenamefont {Burke}, \citenamefont
  {Hoffmann}, \citenamefont {Fahlvik}, \citenamefont {Thelander}, \citenamefont
  {Leijnse},\ and\ \citenamefont {Linke}}]{josefsson2018}%
  \BibitemOpen
  \bibfield  {author} {\bibinfo {author} {\bibfnamefont {M.}~\bibnamefont
  {Josefsson}} \textit{et al.,} }\bibfield  {title} {\bibinfo {title} {A
  quantum-dot heat engine operating close to the thermodynamic efficiency
  limits},\ }\href {https://doi.org/10.1038/s41565-018-0200-5} {\bibfield
  {journal} {\bibinfo  {journal} {Nat. Nanotechnol.}\ }\textbf {\bibinfo
  {volume} {13}},\ \bibinfo {pages} {920} (\bibinfo {year} {2018})}\BibitemShut
  {NoStop}%
\bibitem [{\citenamefont {Klatzow}\ \emph {et~al.}(2019)\citenamefont
  {Klatzow}, \citenamefont {Becker}, \citenamefont {Ledingham}, \citenamefont
  {Weinzetl}, \citenamefont {Kaczmarek}, \citenamefont {Saunders},
  \citenamefont {Nunn}, \citenamefont {Walmsley}, \citenamefont {Uzdin},\ and\
  \citenamefont {Poem}}]{klatzow2019}%
  \BibitemOpen
  \bibfield  {author} {\bibinfo {author} {\bibfnamefont {J.}~\bibnamefont
  {Klatzow}} \textit{et al.,} }\bibfield  {title} {\bibinfo {title}
  {Experimental demonstration of quantum effects in the operation of
  microscopic heat engines},\ }\href
  {https://doi.org/10.1103/PhysRevLett.122.110601} {\bibfield  {journal}
  {\bibinfo  {journal} {Phys. Rev. Lett.}\ }\textbf {\bibinfo {volume} {122}},\
  \bibinfo {pages} {110601} (\bibinfo {year} {2019})}\BibitemShut {NoStop}%
\bibitem [{\citenamefont {von Lindenfels}\ \emph {et~al.}(2019)\citenamefont
  {von Lindenfels}, \citenamefont {Gr\"ab}, \citenamefont {Schmiegelow},
  \citenamefont {Kaushal}, \citenamefont {Schulz}, \citenamefont {Mitchison},
  \citenamefont {Goold}, \citenamefont {Schmidt-Kaler},\ and\ \citenamefont
  {Poschinger}}]{lindenfels2019}%
  \BibitemOpen
  \bibfield  {author} {\bibinfo {author} {\bibfnamefont {D.}~\bibnamefont {von
  Lindenfels}} \textit{et al.,} }\bibfield  {title} {\bibinfo {title}
  {Spin heat engine coupled to a harmonic-oscillator flywheel},\ }\href
  {https://doi.org/10.1103/PhysRevLett.123.080602} {\bibfield  {journal}
  {\bibinfo  {journal} {Phys. Rev. Lett.}\ }\textbf {\bibinfo {volume} {123}},\
  \bibinfo {pages} {080602} (\bibinfo {year} {2019})}\BibitemShut {NoStop}%
\bibitem [{\citenamefont {Maslennikov}\ \emph {et~al.}(2019)\citenamefont
  {Maslennikov}, \citenamefont {Ding}, \citenamefont {Habl{\"a}tzel},
  \citenamefont {Gan}, \citenamefont {Roulet}, \citenamefont {Nimmrichter},
  \citenamefont {Dai}, \citenamefont {Scarani},\ and\ \citenamefont
  {Matsukevich}}]{maslennikov2019}%
  \BibitemOpen
  \bibfield  {author} {\bibinfo {author} {\bibfnamefont {G.}~\bibnamefont
  {Maslennikov}}, \textit{et al.,} }\bibfield  {title} {\bibinfo {title} {Quantum absorption
  refrigerator with trapped ions},\ }\href
  {https://doi.org/10.1038/s41467-018-08090-0} {\bibfield  {journal} {\bibinfo
  {journal} {Nat. Commun.}\ }\textbf {\bibinfo {volume} {10}},\ \bibinfo
  {pages} {202} (\bibinfo {year} {2019})}\BibitemShut {NoStop}%
\bibitem [{\citenamefont {Peterson}\ \emph {et~al.}(2019)\citenamefont
  {Peterson}, \citenamefont {Batalh\~ao}, \citenamefont {Herrera},
  \citenamefont {Souza}, \citenamefont {Sarthour}, \citenamefont {Oliveira},\
  and\ \citenamefont {Serra}}]{peterson2019}%
  \BibitemOpen
  \bibfield  {author} {\bibinfo {author} {\bibfnamefont {J.~P.~S.}\
  \bibnamefont {Peterson}} \textit{et al.,} }\bibfield  {title} {\bibinfo {title} {Experimental
  characterization of a spin quantum heat engine},\ }\href
  {https://doi.org/10.1103/PhysRevLett.123.240601} {\bibfield  {journal}
  {\bibinfo  {journal} {Phys. Rev. Lett.}\ }\textbf {\bibinfo {volume} {123}},\
  \bibinfo {pages} {240601} (\bibinfo {year} {2019})}\BibitemShut {NoStop}%
\bibitem [{\citenamefont {Prete}\ \emph {et~al.}(2019)\citenamefont {Prete},
  \citenamefont {Erdman}, \citenamefont {Demontis}, \citenamefont {Zannier},
  \citenamefont {Ercolani}, \citenamefont {Sorba}, \citenamefont {Beltram},
  \citenamefont {Rossella}, \citenamefont {Taddei},\ and\ \citenamefont
  {Roddaro}}]{prete2019}%
  \BibitemOpen
  \bibfield  {author} {\bibinfo {author} {\bibfnamefont {D.}~\bibnamefont
  {Prete}} \textit{et al.,} }\bibfield  {title} {\bibinfo {title}
  {Thermoelectric conversion at 30 k in inas/inp nanowire quantum dots},\
  }\href {https://doi.org/10.1021/acs.nanolett.9b00276} {\bibfield  {journal}
  {\bibinfo  {journal} {Nano Lett.}\ }\textbf {\bibinfo {volume} {19}},\
  \bibinfo {pages} {3033} (\bibinfo {year} {2019})}\BibitemShut {NoStop}%
\bibitem [{\citenamefont {Horne}\ \emph {et~al.}(2020)\citenamefont {Horne},
  \citenamefont {Yum}, \citenamefont {Dutta}, \citenamefont {H{\"a}nggi},
  \citenamefont {Gong}, \citenamefont {Poletti},\ and\ \citenamefont
  {Mukherjee}}]{horne2020}%
  \BibitemOpen
  \bibfield  {author} {\bibinfo {author} {\bibfnamefont {N.~V.}\ \bibnamefont
  {Horne}} \textit{et al.,} }\bibfield  {title} {\bibinfo {title}
  {Single-atom energy-conversion device with a quantum load},\ }\href
  {https://doi.org/10.1038/s41534-020-0264-6} {\bibfield  {journal} {\bibinfo
  {journal} {NPJ Quantum Inf.}\ }\textbf {\bibinfo {volume} {6}},\ \bibinfo
  {pages} {37} (\bibinfo {year} {2020})}\BibitemShut {NoStop}%
\bibitem [{\citenamefont {Alicki}(1979)}]{alicki1979}%
  \BibitemOpen
  \bibfield  {author} {\bibinfo {author} {\bibfnamefont {R.}~\bibnamefont
  {Alicki}},\ }\bibfield  {title} {\bibinfo {title} {The quantum open system as
  a model of the heat engine},\ }\href
  {https://doi.org/10.1088/0305-4470/12/5/007} {\bibfield  {journal} {\bibinfo
  {journal} {J. Phys. A: Math. Gen.}\ }\textbf {\bibinfo {volume} {12}},\
  \bibinfo {pages} {L103} (\bibinfo {year} {1979})}\BibitemShut {NoStop}%
\bibitem [{\citenamefont {Esposito}\ \emph
  {et~al.}(2010{\natexlab{a}})\citenamefont {Esposito}, \citenamefont {Kawai},
  \citenamefont {Lindenberg},\ and\ \citenamefont {den
  Broeck}}]{esposito2010_prl}%
  \BibitemOpen
  \bibfield  {author} {\bibinfo {author} {\bibfnamefont {M.}~\bibnamefont
  {Esposito}}, \bibinfo {author} {\bibfnamefont {R.}~\bibnamefont {Kawai}},
  \bibinfo {author} {\bibfnamefont {K.}~\bibnamefont {Lindenberg}},\ and\
  \bibinfo {author} {\bibfnamefont {C.~V.}\ \bibnamefont {den Broeck}},\
  }\bibfield  {title} {\bibinfo {title} {Efficiency at maximum power of
  low-dissipation carnot engines},\ }\href
  {https://doi.org/10.1103/PhysRevLett.105.150603} {\bibfield  {journal}
  {\bibinfo  {journal} {Phys. Rev. Lett.}\ }\textbf {\bibinfo {volume} {105}},\
  \bibinfo {pages} {150603} (\bibinfo {year} {2010}{\natexlab{a}})}\BibitemShut
  {NoStop}%
\bibitem [{\citenamefont {Wang}\ \emph {et~al.}(2011)\citenamefont {Wang},
  \citenamefont {He},\ and\ \citenamefont {He}}]{wang2011}%
  \BibitemOpen
  \bibfield  {author} {\bibinfo {author} {\bibfnamefont {J.}~\bibnamefont
  {Wang}}, \bibinfo {author} {\bibfnamefont {J.}~\bibnamefont {He}},\ and\
  \bibinfo {author} {\bibfnamefont {X.}~\bibnamefont {He}},\ }\bibfield
  {title} {\bibinfo {title} {Performance analysis of a two-state quantum heat
  engine working with a single-mode radiation field in a cavity},\ }\href
  {https://doi.org/10.1103/PhysRevE.84.041127} {\bibfield  {journal} {\bibinfo
  {journal} {Phys. Rev. E}\ }\textbf {\bibinfo {volume} {84}},\ \bibinfo
  {pages} {041127} (\bibinfo {year} {2011})}\BibitemShut {NoStop}%
\bibitem [{\citenamefont {Avron}\ \emph {et~al.}(2012)\citenamefont {Avron},
  \citenamefont {Fraas}, \citenamefont {Graf},\ and\ \citenamefont
  {Grech}}]{avron2012}%
  \BibitemOpen
  \bibfield  {author} {\bibinfo {author} {\bibfnamefont {J.~E.}\ \bibnamefont
  {Avron}}, \bibinfo {author} {\bibfnamefont {M.}~\bibnamefont {Fraas}},
  \bibinfo {author} {\bibfnamefont {G.~M.}\ \bibnamefont {Graf}},\ and\
  \bibinfo {author} {\bibfnamefont {P.}~\bibnamefont {Grech}},\ }\bibfield
  {title} {\bibinfo {title} {Adiabatic theorems for generators of contracting
  evolutions},\ }\href {https://doi.org/10.1007/s00220-012-1504-1} {\bibfield
  {journal} {\bibinfo  {journal} {Commun. Math. Phys.}\ }\textbf {\bibinfo
  {volume} {314}},\ \bibinfo {pages} {163} (\bibinfo {year}
  {2012})}\BibitemShut {NoStop}%
\bibitem [{\citenamefont {Ludovico}\ \emph {et~al.}(2016)\citenamefont
  {Ludovico}, \citenamefont {Battista}, \citenamefont {von Oppen},\ and\
  \citenamefont {Arrachea}}]{ludovico2016}%
  \BibitemOpen
  \bibfield  {author} {\bibinfo {author} {\bibfnamefont {M.~F.}\ \bibnamefont
  {Ludovico}}, \bibinfo {author} {\bibfnamefont {F.}~\bibnamefont {Battista}},
  \bibinfo {author} {\bibfnamefont {F.}~\bibnamefont {von Oppen}},\ and\
  \bibinfo {author} {\bibfnamefont {L.}~\bibnamefont {Arrachea}},\ }\bibfield
  {title} {\bibinfo {title} {Adiabatic response and quantum thermoelectrics for
  ac-driven quantum systems},\ }\href
  {https://doi.org/10.1103/PhysRevB.93.075136} {\bibfield  {journal} {\bibinfo
  {journal} {Phys. Rev. B}\ }\textbf {\bibinfo {volume} {93}},\ \bibinfo
  {pages} {075136} (\bibinfo {year} {2016})}\BibitemShut {NoStop}%
\bibitem [{\citenamefont {Cavina}\ \emph {et~al.}(2017)\citenamefont {Cavina},
  \citenamefont {Mari},\ and\ \citenamefont {Giovannetti}}]{cavina2017_prl}%
  \BibitemOpen
  \bibfield  {author} {\bibinfo {author} {\bibfnamefont {V.}~\bibnamefont
  {Cavina}}, \bibinfo {author} {\bibfnamefont {A.}~\bibnamefont {Mari}},\ and\
  \bibinfo {author} {\bibfnamefont {V.}~\bibnamefont {Giovannetti}},\
  }\bibfield  {title} {\bibinfo {title} {Slow dynamics and thermodynamics of
  open quantum systems},\ }\href
  {https://doi.org/10.1103/PhysRevLett.119.050601} {\bibfield  {journal}
  {\bibinfo  {journal} {Phys. Rev. Lett.}\ }\textbf {\bibinfo {volume} {119}},\
  \bibinfo {pages} {050601} (\bibinfo {year} {2017})}\BibitemShut {NoStop}%
\bibitem [{\citenamefont {Abiuso}\ and\ \citenamefont
  {Giovannetti}(2019)}]{abiuso2018}%
  \BibitemOpen
  \bibfield  {author} {\bibinfo {author} {\bibfnamefont {P.}~\bibnamefont
  {Abiuso}}\ and\ \bibinfo {author} {\bibfnamefont {V.}~\bibnamefont
  {Giovannetti}},\ }\bibfield  {title} {\bibinfo {title} {Non-markov
  enhancement of maximum power for quantum thermal machines},\ }\href
  {https://doi.org/10.1103/PhysRevA.99.052106} {\bibfield  {journal} {\bibinfo
  {journal} {Phys. Rev. A}\ }\textbf {\bibinfo {volume} {99}},\ \bibinfo
  {pages} {052106} (\bibinfo {year} {2019})}\BibitemShut {NoStop}%
\bibitem [{\citenamefont {Bhandari}\ \emph {et~al.}(2020)\citenamefont
  {Bhandari}, \citenamefont {Alonso}, \citenamefont {Taddei}, \citenamefont
  {von Oppen}, \citenamefont {Fazio},\ and\ \citenamefont
  {Arrachea}}]{bhandari2020}%
  \BibitemOpen
  \bibfield  {author} {\bibinfo {author} {\bibfnamefont {B.}~\bibnamefont
  {Bhandari}} \textit{et al.,} }\bibfield  {title} {\bibinfo
  {title} {Geometric properties of adiabatic quantum thermal machines},\
  }\href{https://doi.org/10.1103/PhysRevB.102.155407} {\bibfield  {journal}
  {\bibinfo  {journal} {Phys. Rev. B}\ }\textbf {\bibinfo {volume} {102}},\
  \bibinfo {pages} {155407} (\bibinfo {year} {2020})}\BibitemShut
  {NoStop}%
\bibitem [{\citenamefont {Abiuso}\ and\ \citenamefont
  {Perarnau-Llobet}(2020)}]{abiuso2020_prl}%
  \BibitemOpen
  \bibfield  {author} {\bibinfo {author} {\bibfnamefont {P.}~\bibnamefont
  {Abiuso}}\ and\ \bibinfo {author} {\bibfnamefont {M.}~\bibnamefont
  {Perarnau-Llobet}},\ }\bibfield  {title} {\bibinfo {title} {Optimal cycles
  for low-dissipation heat engines},\ }\href
  {https://doi.org/10.1103/PhysRevLett.124.110606} {\bibfield  {journal}
  {\bibinfo  {journal} {Phys. Rev. Lett.}\ }\textbf {\bibinfo {volume} {124}},\
  \bibinfo {pages} {110606} (\bibinfo {year} {2020})}\BibitemShut {NoStop}%
\bibitem [{\citenamefont {Abiuso}\ \emph {et~al.}(2020)\citenamefont {Abiuso},
  \citenamefont {Mille}, \citenamefont {Perarnau-Llobet},\ and\ \citenamefont
  {Scandi}}]{abiuso2020_entropy}%
  \BibitemOpen
  \bibfield  {author} {\bibinfo {author} {\bibfnamefont {P.}~\bibnamefont
  {Abiuso}}, \bibinfo {author} {\bibfnamefont {H.~J.~D.}\ \bibnamefont
  {Mille}}, \bibinfo {author} {\bibfnamefont {M.}~\bibnamefont
  {Perarnau-Llobet}},\ and\ \bibinfo {author} {\bibfnamefont {M.}~\bibnamefont
  {Scandi}},\ }\bibfield  {title} {\bibinfo {title} {Geometric optimisation of
  quantum thermodynamic processes},\ }\href {https://doi.org/10.3390/e22101076}
  {\bibfield  {journal} {\bibinfo  {journal} {Entropy}\ }\textbf {\bibinfo
  {volume} {22}} (\bibinfo {year} {2020})}\BibitemShut {NoStop}%
\bibitem [{\citenamefont {Cavina}\ \emph {et~al.}(2020)\citenamefont {Cavina},
  \citenamefont {Erdman}, \citenamefont {Abiuso}, \citenamefont {Tolomeo},\
  and\ \citenamefont {Giovannetti}}]{cavina2020}%
  \BibitemOpen
  \bibfield  {author} {\bibinfo {author} {\bibfnamefont {V.}~\bibnamefont
  {Cavina}}, \bibinfo {author} {\bibfnamefont {P.~A.}\ \bibnamefont {Erdman}},
  \bibinfo {author} {\bibfnamefont {P.}~\bibnamefont {Abiuso}}, \bibinfo
  {author} {\bibfnamefont {L.}~\bibnamefont {Tolomeo}},\ and\ \bibinfo {author}
  {\bibfnamefont {V.}~\bibnamefont {Giovannetti}},\ }\bibfield  {title}
  {\bibinfo {title} {Maximum power heat engines and refrigerators in the
  fast-driving regime},\ }\href{https://doi.org/10.1103/PhysRevA.104.032226} {\bibfield  {journal} {\bibinfo
  {journal} {Phys. Rev. A}\ }\textbf {\bibinfo {volume} {104}},\ \bibinfo
  {pages} {032226} (\bibinfo {year} {2021})}\BibitemShut {NoStop}%
\bibitem [{\citenamefont {Arrachea}\ \emph {et~al.}(2007)\citenamefont
  {Arrachea}, \citenamefont {Moskalets},\ and\ \citenamefont
  {Martin-Moreno}}]{arrachea2007}%
  \BibitemOpen
  \bibfield  {author} {\bibinfo {author} {\bibfnamefont {L.}~\bibnamefont
  {Arrachea}}, \bibinfo {author} {\bibfnamefont {M.}~\bibnamefont
  {Moskalets}},\ and\ \bibinfo {author} {\bibfnamefont {L.}~\bibnamefont
  {Martin-Moreno}},\ }\bibfield  {title} {\bibinfo {title} {Heat production and
  energy balance in nanoscale engines driven by time-dependent fields},\ }\href
  {https://doi.org/10.1103/PhysRevB.75.245420} {\bibfield  {journal} {\bibinfo
  {journal} {Phys. Rev. B}\ }\textbf {\bibinfo {volume} {75}},\ \bibinfo
  {pages} {245420} (\bibinfo {year} {2007})}\BibitemShut {NoStop}%
\bibitem [{\citenamefont {Esposito}\ \emph
  {et~al.}(2010{\natexlab{b}})\citenamefont {Esposito}, \citenamefont {Kawai},
  \citenamefont {Lindenberg},\ and\ \citenamefont {Van~den
  Broeck}}]{esposito2010_pre}%
  \BibitemOpen
  \bibfield  {author} {\bibinfo {author} {\bibfnamefont {M.}~\bibnamefont
  {Esposito}}, \bibinfo {author} {\bibfnamefont {R.}~\bibnamefont {Kawai}},
  \bibinfo {author} {\bibfnamefont {K.}~\bibnamefont {Lindenberg}},\ and\
  \bibinfo {author} {\bibfnamefont {C.}~\bibnamefont {Van~den Broeck}},\
  }\bibfield  {title} {\bibinfo {title} {Quantum-dot carnot engine at maximum
  power},\ }\href {https://doi.org/10.1103/PhysRevE.81.041106} {\bibfield
  {journal} {\bibinfo  {journal} {Phys. Rev. E}\ }\textbf {\bibinfo {volume}
  {81}},\ \bibinfo {pages} {041106} (\bibinfo {year}
  {2010}{\natexlab{b}})}\BibitemShut {NoStop}%
\bibitem [{\citenamefont {Juergens}\ \emph {et~al.}(2013)\citenamefont
  {Juergens}, \citenamefont {Haupt}, \citenamefont {Moskalets},\ and\
  \citenamefont {Splettstoesser}}]{juergens2013}%
  \BibitemOpen
  \bibfield  {author} {\bibinfo {author} {\bibfnamefont {S.}~\bibnamefont
  {Juergens}}, \bibinfo {author} {\bibfnamefont {F.}~\bibnamefont {Haupt}},
  \bibinfo {author} {\bibfnamefont {M.}~\bibnamefont {Moskalets}},\ and\
  \bibinfo {author} {\bibfnamefont {J.}~\bibnamefont {Splettstoesser}},\
  }\bibfield  {title} {\bibinfo {title} {Thermoelectric performance of a driven
  double quantum dot},\ }\href {https://doi.org/10.1103/PhysRevB.87.245423}
  {\bibfield  {journal} {\bibinfo  {journal} {Phys. Rev. B}\ }\textbf {\bibinfo
  {volume} {87}},\ \bibinfo {pages} {245423} (\bibinfo {year}
  {2013})}\BibitemShut {NoStop}%
\bibitem [{\citenamefont {Campisi}\ \emph {et~al.}(2015)\citenamefont
  {Campisi}, \citenamefont {Pekola},\ and\ \citenamefont
  {Fazio}}]{campisi2015}%
  \BibitemOpen
  \bibfield  {author} {\bibinfo {author} {\bibfnamefont {M.}~\bibnamefont
  {Campisi}}, \bibinfo {author} {\bibfnamefont {J.}~\bibnamefont {Pekola}},\
  and\ \bibinfo {author} {\bibfnamefont {R.}~\bibnamefont {Fazio}},\ }\bibfield
   {title} {\bibinfo {title} {Nonequilibrium fluctuations in quantum heat
  engines: theory, example, and possible solid state experiments},\ }\href
  {https://doi.org/10.1088/1367-2630/17/3/035012} {\bibfield  {journal}
  {\bibinfo  {journal} {New J. Phys.}\ }\textbf {\bibinfo {volume} {17}},\
  \bibinfo {pages} {035012} (\bibinfo {year} {2015})}\BibitemShut {NoStop}%
\bibitem [{\citenamefont {Dann}\ and\ \citenamefont
  {Kosloff}(2020)}]{dann2020}%
  \BibitemOpen
  \bibfield  {author} {\bibinfo {author} {\bibfnamefont {R.}~\bibnamefont
  {Dann}}\ and\ \bibinfo {author} {\bibfnamefont {R.}~\bibnamefont {Kosloff}},\
  }\bibfield  {title} {\bibinfo {title} {Quantum signatures in the quantum
  carnot cycle},\ }\href {https://doi.org/10.1088/1367-2630/ab6876} {\bibfield
  {journal} {\bibinfo  {journal} {New J. Phys.}\ }\textbf {\bibinfo {volume}
  {22}},\ \bibinfo {pages} {013055} (\bibinfo {year} {2020})}\BibitemShut
  {NoStop}%
\bibitem [{\citenamefont {Molitor}\ and\ \citenamefont
  {Landi}(2020)}]{molitor2020}%
  \BibitemOpen
  \bibfield  {author} {\bibinfo {author} {\bibfnamefont {O.~A.~D.}\
  \bibnamefont {Molitor}}\ and\ \bibinfo {author} {\bibfnamefont {G.~T.}\
  \bibnamefont {Landi}},\ }\bibfield  {title} {\bibinfo {title} {Stroboscopic
  two-stroke quantum heat engines},\ }\href
  {https://doi.org/10.1103/PhysRevA.102.042217} {\bibfield  {journal} {\bibinfo
   {journal} {Phys. Rev. A}\ }\textbf {\bibinfo {volume} {102}},\ \bibinfo
  {pages} {042217} (\bibinfo {year} {2020})}\BibitemShut {NoStop}%
\bibitem [{\citenamefont {Feldmann}\ \emph {et~al.}(1996)\citenamefont
  {Feldmann}, \citenamefont {Geva}, \citenamefont {Kosloff},\ and\
  \citenamefont {Salamon}}]{feldmann1996}%
  \BibitemOpen
  \bibfield  {author} {\bibinfo {author} {\bibfnamefont {T.}~\bibnamefont
  {Feldmann}}, \bibinfo {author} {\bibfnamefont {E.}~\bibnamefont {Geva}},
  \bibinfo {author} {\bibfnamefont {R.}~\bibnamefont {Kosloff}},\ and\ \bibinfo
  {author} {\bibfnamefont {P.}~\bibnamefont {Salamon}},\ }\bibfield  {title}
  {\bibinfo {title} {Heat engines in finite time governed by master
  equations},\ }\href {https://doi.org/10.1119/1.18197} {\bibfield  {journal}
  {\bibinfo  {journal} {Am. J. Phys.}\ }\textbf {\bibinfo {volume} {64}},\
  \bibinfo {pages} {485} (\bibinfo {year} {1996})}\BibitemShut {NoStop}%
\bibitem [{\citenamefont {Feldmann}\ and\ \citenamefont
  {Kosloff}(2000)}]{feldmann2000}%
  \BibitemOpen
  \bibfield  {author} {\bibinfo {author} {\bibfnamefont {T.}~\bibnamefont
  {Feldmann}}\ and\ \bibinfo {author} {\bibfnamefont {R.}~\bibnamefont
  {Kosloff}},\ }\bibfield  {title} {\bibinfo {title} {Performance of discrete
  heat engines and heat pumps in finite time},\ }\href
  {https://doi.org/10.1103/PhysRevE.61.4774} {\bibfield  {journal} {\bibinfo
  {journal} {Phys. Rev. E}\ }\textbf {\bibinfo {volume} {61}},\ \bibinfo
  {pages} {4774} (\bibinfo {year} {2000})}\BibitemShut {NoStop}%
\bibitem [{\citenamefont {Rezek}\ and\ \citenamefont
  {Kosloff}(2006)}]{rezek2006}%
  \BibitemOpen
  \bibfield  {author} {\bibinfo {author} {\bibfnamefont {Y.}~\bibnamefont
  {Rezek}}\ and\ \bibinfo {author} {\bibfnamefont {R.}~\bibnamefont
  {Kosloff}},\ }\bibfield  {title} {\bibinfo {title} {Irreversible performance
  of a quantum harmonic heat engine},\ }\href
  {https://doi.org/10.1088/1367-2630/8/5/083} {\bibfield  {journal} {\bibinfo
  {journal} {New J. Phys.}\ }\textbf {\bibinfo {volume} {8}},\ \bibinfo {pages}
  {83} (\bibinfo {year} {2006})}\BibitemShut {NoStop}%
\bibitem [{\citenamefont {Quan}\ \emph {et~al.}(2007)\citenamefont {Quan},
  \citenamefont {Liu}, \citenamefont {Sun},\ and\ \citenamefont
  {Nori}}]{quan2007}%
  \BibitemOpen
  \bibfield  {author} {\bibinfo {author} {\bibfnamefont {H.}~\bibnamefont
  {Quan}}, \bibinfo {author} {\bibfnamefont {Y.}~\bibnamefont {Liu}}, \bibinfo
  {author} {\bibfnamefont {C.}~\bibnamefont {Sun}},\ and\ \bibinfo {author}
  {\bibfnamefont {F.}~\bibnamefont {Nori}},\ }\bibfield  {title} {\bibinfo
  {title} {Quantum thermodynamic cycles and quantum heat engines},\ }\href
  {https://doi.org/10.1103/PhysRevE.76.031105} {\bibfield  {journal} {\bibinfo
  {journal} {Phys. Rev. E}\ }\textbf {\bibinfo {volume} {76}},\ \bibinfo
  {pages} {031105} (\bibinfo {year} {2007})}\BibitemShut {NoStop}%
\bibitem [{\citenamefont {Abah}\ \emph {et~al.}(2012)\citenamefont {Abah},
  \citenamefont {Ro\ss{}nagel}, \citenamefont {Jacob}, \citenamefont {Deffner},
  \citenamefont {Schmidt-Kaler}, \citenamefont {Singer},\ and\ \citenamefont
  {Lutz}}]{abah2012}%
  \BibitemOpen
  \bibfield  {author} {\bibinfo {author} {\bibfnamefont {O.}~\bibnamefont
  {Abah}} \textit{et al.,} }\bibfield  {title} {\bibinfo
  {title} {Single-ion heat engine at maximum power},\ }\href
  {https://doi.org/10.1103/PhysRevLett.109.203006} {\bibfield  {journal}
  {\bibinfo  {journal} {Phys. Rev. Lett.}\ }\textbf {\bibinfo {volume} {109}},\
  \bibinfo {pages} {203006} (\bibinfo {year} {2012})}\BibitemShut {NoStop}%
\bibitem [{\citenamefont {Allahverdyan}\ \emph {et~al.}(2013)\citenamefont
  {Allahverdyan}, \citenamefont {Hovhannisyan}, \citenamefont {Melkikh},\ and\
  \citenamefont {Gevorkian}}]{allahverdyan2013}%
  \BibitemOpen
  \bibfield  {author} {\bibinfo {author} {\bibfnamefont {A.~E.}\ \bibnamefont
  {Allahverdyan}}, \bibinfo {author} {\bibfnamefont {K.~V.}\ \bibnamefont
  {Hovhannisyan}}, \bibinfo {author} {\bibfnamefont {A.~V.}\ \bibnamefont
  {Melkikh}},\ and\ \bibinfo {author} {\bibfnamefont {S.~G.}\ \bibnamefont
  {Gevorkian}},\ }\bibfield  {title} {\bibinfo {title} {Carnot cycle at finite
  power: Attainability of maximal efficiency},\ }\href
  {https://doi.org/10.1103/PhysRevLett.111.050601} {\bibfield  {journal}
  {\bibinfo  {journal} {Phys. Rev. Lett.}\ }\textbf {\bibinfo {volume} {111}},\
  \bibinfo {pages} {050601} (\bibinfo {year} {2013})}\BibitemShut {NoStop}%
\bibitem [{\citenamefont {Zhang}\ \emph {et~al.}(2014)\citenamefont {Zhang},
  \citenamefont {Bariani},\ and\ \citenamefont {Meystre}}]{zhang2014}%
  \BibitemOpen
  \bibfield  {author} {\bibinfo {author} {\bibfnamefont {K.}~\bibnamefont
  {Zhang}}, \bibinfo {author} {\bibfnamefont {F.}~\bibnamefont {Bariani}},\
  and\ \bibinfo {author} {\bibfnamefont {P.}~\bibnamefont {Meystre}},\
  }\bibfield  {title} {\bibinfo {title} {Quantum optomechanical heat engine},\
  }\href {https://doi.org/10.1103/PhysRevLett.112.150602} {\bibfield  {journal}
  {\bibinfo  {journal} {Phys. Rev. Lett.}\ }\textbf {\bibinfo {volume} {112}},\
  \bibinfo {pages} {150602} (\bibinfo {year} {2014})}\BibitemShut {NoStop}%
\bibitem [{\citenamefont {Campisi}\ and\ \citenamefont
  {Fazio}(2016)}]{campisi2016}%
  \BibitemOpen
  \bibfield  {author} {\bibinfo {author} {\bibfnamefont {M.}~\bibnamefont
  {Campisi}}\ and\ \bibinfo {author} {\bibfnamefont {R.}~\bibnamefont
  {Fazio}},\ }\bibfield  {title} {\bibinfo {title} {The power of a critical
  heat engine},\ }\href {https://doi.org/10.1038/ncomms11895} {\bibfield
  {journal} {\bibinfo  {journal} {Nat. Commun.}\ }\textbf {\bibinfo {volume}
  {7}},\ \bibinfo {pages} {11895} (\bibinfo {year} {2016})}\BibitemShut
  {NoStop}%
\bibitem [{\citenamefont {Karimi}\ and\ \citenamefont
  {Pekola}(2016)}]{karimi2016}%
  \BibitemOpen
  \bibfield  {author} {\bibinfo {author} {\bibfnamefont {B.}~\bibnamefont
  {Karimi}}\ and\ \bibinfo {author} {\bibfnamefont {J.~P.}\ \bibnamefont
  {Pekola}},\ }\bibfield  {title} {\bibinfo {title} {Otto refrigerator based on
  a superconducting qubit: Classical and quantum performance},\ }\href
  {https://doi.org/10.1103/PhysRevB.94.184503} {\bibfield  {journal} {\bibinfo
  {journal} {Phys. Rev. B}\ }\textbf {\bibinfo {volume} {94}},\ \bibinfo
  {pages} {184503} (\bibinfo {year} {2016})}\BibitemShut {NoStop}%
\bibitem [{\citenamefont {Kosloff}\ and\ \citenamefont
  {Rezek}(2017)}]{kosloff2017}%
  \BibitemOpen
  \bibfield  {author} {\bibinfo {author} {\bibfnamefont {R.}~\bibnamefont
  {Kosloff}}\ and\ \bibinfo {author} {\bibfnamefont {Y.}~\bibnamefont
  {Rezek}},\ }\bibfield  {title} {\bibinfo {title} {The quantum harmonic otto
  cycle},\ }\href {https://doi.org/10.3390/e19040136} {\bibfield  {journal}
  {\bibinfo  {journal} {Entropy}\ }\textbf {\bibinfo {volume} {19}},\ \bibinfo
  {pages} {136} (\bibinfo {year} {2017})}\BibitemShut {NoStop}%
\bibitem [{\citenamefont {Watanabe}\ \emph {et~al.}(2017)\citenamefont
  {Watanabe}, \citenamefont {Venkatesh}, \citenamefont {Talkner},\ and\
  \citenamefont {del Campo}}]{watanabe2017}%
  \BibitemOpen
  \bibfield  {author} {\bibinfo {author} {\bibfnamefont {G.}~\bibnamefont
  {Watanabe}}, \bibinfo {author} {\bibfnamefont {B.~P.}\ \bibnamefont
  {Venkatesh}}, \bibinfo {author} {\bibfnamefont {P.}~\bibnamefont {Talkner}},\
  and\ \bibinfo {author} {\bibfnamefont {A.}~\bibnamefont {del Campo}},\
  }\bibfield  {title} {\bibinfo {title} {Quantum performance of thermal
  machines over many cycles},\ }\href
  {https://doi.org/10.1103/PhysRevLett.118.050601} {\bibfield  {journal}
  {\bibinfo  {journal} {Phys. Rev. Lett.}\ }\textbf {\bibinfo {volume} {118}},\
  \bibinfo {pages} {050601} (\bibinfo {year} {2017})}\BibitemShut {NoStop}%
\bibitem [{\citenamefont {Deffner}(2018)}]{deffner2018}%
  \BibitemOpen
  \bibfield  {author} {\bibinfo {author} {\bibfnamefont {S.}~\bibnamefont
  {Deffner}},\ }\bibfield  {title} {\bibinfo {title} {Efficiency of harmonic
  quantum otto engines at maximal power},\ }\href
  {https://doi.org/10.3390/e20110875} {\bibfield  {journal} {\bibinfo
  {journal} {Entropy}\ }\textbf {\bibinfo {volume} {20}},\ \bibinfo {pages}
  {875} (\bibinfo {year} {2018})}\BibitemShut {NoStop}%
\bibitem [{\citenamefont {Gelbwaser-Klimovsky}\ \emph
  {et~al.}(2018)\citenamefont {Gelbwaser-Klimovsky}, \citenamefont {Bylinskii},
  \citenamefont {Gangloff}, \citenamefont {Islam}, \citenamefont
  {Aspuru-Guzik},\ and\ \citenamefont {Vuletic}}]{gelbwaser2018}%
  \BibitemOpen
  \bibfield  {author} {\bibinfo {author} {\bibfnamefont {D.}~\bibnamefont
  {Gelbwaser-Klimovsky}} \textit{et al.,} }\bibfield  {title}
  {\bibinfo {title} {Single-atom heat machines enabled by energy
  quantization},\ }\href {https://doi.org/10.1103/PhysRevLett.120.170601}
  {\bibfield  {journal} {\bibinfo  {journal} {Phys. Rev. Lett.}\ }\textbf
  {\bibinfo {volume} {120}},\ \bibinfo {pages} {170601} (\bibinfo {year}
  {2018})}\BibitemShut {NoStop}%
\bibitem [{\citenamefont {Chen}\ \emph {et~al.}(2019)\citenamefont {Chen},
  \citenamefont {Sun},\ and\ \citenamefont {Dong}}]{chen2019}%
  \BibitemOpen
  \bibfield  {author} {\bibinfo {author} {\bibfnamefont {J.}~\bibnamefont
  {Chen}}, \bibinfo {author} {\bibfnamefont {C.}~\bibnamefont {Sun}},\ and\
  \bibinfo {author} {\bibfnamefont {H.}~\bibnamefont {Dong}},\ }\bibfield
  {title} {\bibinfo {title} {Boosting the performance of quantum otto heat
  engines},\ }\href {https://doi.org/10.1103/PhysRevE.100.032144} {\bibfield
  {journal} {\bibinfo  {journal} {Phys. Rev. E}\ }\textbf {\bibinfo {volume}
  {100}},\ \bibinfo {pages} {032144} (\bibinfo {year} {2019})}\BibitemShut
  {NoStop}%
\bibitem [{\citenamefont {Pekola}\ \emph {et~al.}(2019)\citenamefont {Pekola},
  \citenamefont {Karimi}, \citenamefont {Thomas},\ and\ \citenamefont
  {Averin}}]{pekola2019}%
  \BibitemOpen
  \bibfield  {author} {\bibinfo {author} {\bibfnamefont {J.~P.}\ \bibnamefont
  {Pekola}}, \bibinfo {author} {\bibfnamefont {B.}~\bibnamefont {Karimi}},
  \bibinfo {author} {\bibfnamefont {G.}~\bibnamefont {Thomas}},\ and\ \bibinfo
  {author} {\bibfnamefont {D.~V.}\ \bibnamefont {Averin}},\ }\bibfield  {title}
  {\bibinfo {title} {Supremacy of incoherent sudden cycles},\ }\href
  {https://doi.org/10.1103/PhysRevB.100.085405} {\bibfield  {journal} {\bibinfo
   {journal} {Phys. Rev. B}\ }\textbf {\bibinfo {volume} {100}},\ \bibinfo
  {pages} {085405} (\bibinfo {year} {2019})}\BibitemShut {NoStop}%
\bibitem [{\citenamefont {Das}\ and\ \citenamefont
  {Mukherjee}(2020)}]{das2020}%
  \BibitemOpen
  \bibfield  {author} {\bibinfo {author} {\bibfnamefont {A.}~\bibnamefont
  {Das}}\ and\ \bibinfo {author} {\bibfnamefont {V.}~\bibnamefont
  {Mukherjee}},\ }\bibfield  {title} {\bibinfo {title} {Quantum-enhanced
  finite-time otto cycle},\ }\href
  {https://doi.org/10.1103/PhysRevResearch.2.033083} {\bibfield  {journal}
  {\bibinfo  {journal} {Phys. Rev. B}\ }\textbf {\bibinfo {volume} {2}},\
  \bibinfo {pages} {033083} (\bibinfo {year} {2020})}\BibitemShut {NoStop}%
\bibitem [{\citenamefont {Berry}(2009)}]{berry2009}%
  \BibitemOpen
  \bibfield  {author} {\bibinfo {author} {\bibfnamefont {M.~V.}\ \bibnamefont
  {Berry}},\ }\bibfield  {title} {\bibinfo {title} {Transitionless quantum
  driving},\ }\href {https://doi.org/10.1088/1751-8113/42/36/365303} {\bibfield
   {journal} {\bibinfo  {journal} {J. Phys. A: Math. Theor.}\ }\textbf
  {\bibinfo {volume} {42}},\ \bibinfo {pages} {365303} (\bibinfo {year}
  {2009})}\BibitemShut {NoStop}%
\bibitem [{\citenamefont {Deng}\ \emph {et~al.}(2013)\citenamefont {Deng},
  \citenamefont {Liu}, \citenamefont {H{\"a}nggi},\ and\ \citenamefont
  {J.}}]{deng2013}%
  \BibitemOpen
  \bibfield  {author} {\bibinfo {author} {\bibfnamefont {J.}\ \bibnamefont
  {Deng}}, \bibinfo {author} {\bibfnamefont
  {Q.-h.}~\bibnamefont {Wang}}, \bibinfo {author} {\bibfnamefont {Z.}~\bibnamefont
  {Liu}}, \bibinfo {author} {\bibfnamefont {P.}~\bibnamefont
  {H{\"a}nggi}},\ and\  \bibinfo {author} {\bibfnamefont {J.}~\bibnamefont
  {Gong}},\ }\bibfield  {title} {\bibinfo {title} {Boosting work characteristics
  and overall heat-engine performance via shortcuts to adiabaticity: Quantum
  and classical systems},\ }\href {https://doi.org/10.1103/PhysRevE.88.062122}
  {\bibfield  {journal} {\bibinfo  {journal} {Phys. Rev. E}\ }\textbf {\bibinfo
  {volume} {88}},\ \bibinfo {pages} {062122} (\bibinfo {year}
  {2013})}\BibitemShut {NoStop}%
\bibitem [{\citenamefont {Torrontegui}\ \emph {et~al.}(2013)\citenamefont
  {Torrontegui}, \citenamefont {Ib{\'a}{\~n}ez}, \citenamefont
  {Mart{\'\i}nez-Garaot}, \citenamefont {Modugno}, \citenamefont {del Campo},
  \citenamefont {Gu{\'e}ry-Odelin}, \citenamefont {Ruschhaupt}, \citenamefont
  {Chen},\ and\ \citenamefont {Muga}}]{torrontegui2013}%
  \BibitemOpen
  \bibfield  {author} {\bibinfo {author} {\bibfnamefont {E.}~\bibnamefont
  {Torrontegui}} \textit{et al.,} }\bibfield  {title} {\bibinfo
  {title} {Shortcuts to adiabaticity},\ }\href
  {https://doi.org/10.1016/B978-0-12-408090-4.00002-5} {\bibfield  {journal}
  {\bibinfo  {journal} {Adv. At., Mol., Opt. Phys.}\ }\textbf {\bibinfo
  {volume} {62}},\ \bibinfo {pages} {117} (\bibinfo {year} {2013})}\BibitemShut
  {NoStop}%
\bibitem [{\citenamefont {del Campo}\ \emph {et~al.}(2014)\citenamefont {del
  Campo}, \citenamefont {Goold},\ and\ \citenamefont
  {Paternostro}}]{campo2014}%
  \BibitemOpen
  \bibfield  {author} {\bibinfo {author} {\bibfnamefont {A.}~\bibnamefont {del
  Campo}}, \bibinfo {author} {\bibfnamefont {J.}~\bibnamefont {Goold}},\ and\
  \bibinfo {author} {\bibfnamefont {M.}~\bibnamefont {Paternostro}},\
  }\bibfield  {title} {\bibinfo {title} {More bang for your buck:
  Super-adiabatic quantum engines},\ }\href {https://doi.org/10.1038/srep06208}
  {\bibfield  {journal} {\bibinfo  {journal} {Sci. Rep.}\ }\textbf {\bibinfo
  {volume} {4}},\ \bibinfo {pages} {6208} (\bibinfo {year} {2014})}\BibitemShut
  {NoStop}%
\bibitem [{\citenamefont {\c{C}akmak}\ and\ \citenamefont
  {M\"{u}stecapl{\i}o\u{g}lu}(2019)}]{cakmak2018}%
  \BibitemOpen
  \bibfield  {author} {\bibinfo {author} {\bibfnamefont {B.}~\bibnamefont
  {\c{C}akmak}}\ and\ \bibinfo {author} {\bibfnamefont {O.~E.}\ \bibnamefont
  {M\"{u}stecapl{\i}o\u{g}lu}},\ }\bibfield  {title} {\bibinfo {title} {Spin
  quantum heat engines with shortcuts to adiabaticity},\ }\href
  {https://doi.org/10.1103/PhysRevE.99.032108} {\bibfield  {journal} {\bibinfo
  {journal} {Phys. Rev. E}\ }\textbf {\bibinfo {volume} {99}},\ \bibinfo
  {pages} {032108} (\bibinfo {year} {2019})}\BibitemShut {NoStop}%
\bibitem [{\citenamefont {Deng}\ \emph {et~al.}(2018)\citenamefont {Deng},
  \citenamefont {Chenu}, \citenamefont {Diao1}, \citenamefont {Li},
  \citenamefont {Yu}, \citenamefont {Coulamy}, \citenamefont {del Campo},\ and\
  \citenamefont {Wu}}]{deng2018}%
  \BibitemOpen
  \bibfield  {author} {\bibinfo {author} {\bibfnamefont {S.}~\bibnamefont
  {Deng}}  \textit{et al.,} }\bibfield  {title} {\bibinfo {title}
  {Superadiabatic quantum friction suppression in finite-time thermodynamics},\
  }\href {https://doi.org/10.1126/sciadv.aar5909} {\bibfield  {journal}
  {\bibinfo  {journal} {Sci. Adv.}\ }\textbf {\bibinfo {volume} {18}},\
  \bibinfo {pages} {eaar5909} (\bibinfo {year} {2018})}\BibitemShut {NoStop}%
\bibitem [{\citenamefont {Funo}\ \emph {et~al.}(2019)\citenamefont {Funo},
  \citenamefont {Lambert}, \citenamefont {Karimi}, \citenamefont {Pekola},
  \citenamefont {Masuyama},\ and\ \citenamefont {Nori}}]{funo2019}%
  \BibitemOpen
  \bibfield  {author} {\bibinfo {author} {\bibfnamefont {K.}~\bibnamefont
  {Funo}} \textit{et al.,} }\bibfield  {title} {\bibinfo
  {title} {Speeding up a quantum refrigerator via counterdiabatic driving},\
  }\href {https://doi.org/10.1103/PhysRevB.100.035407} {\bibfield  {journal}
  {\bibinfo  {journal} {Phys. Rev. B}\ }\textbf {\bibinfo {volume} {100}},\
  \bibinfo {pages} {035407} (\bibinfo {year} {2019})}\BibitemShut {NoStop}%
\bibitem [{\citenamefont {Villazon}\ \emph {et~al.}(2019)\citenamefont
  {Villazon}, \citenamefont {Polkovnikov},\ and\ \citenamefont
  {Chandran}}]{villazon2019}%
  \BibitemOpen
  \bibfield  {author} {\bibinfo {author} {\bibfnamefont {T.}~\bibnamefont
  {Villazon}}, \bibinfo {author} {\bibfnamefont {A.}~\bibnamefont
  {Polkovnikov}},\ and\ \bibinfo {author} {\bibfnamefont {A.}~\bibnamefont
  {Chandran}},\ }\bibfield  {title} {\bibinfo {title} {Swift heat transfer by
  fast-forward driving in open quantum systems},\ }\href
  {https://doi.org/10.1103/PhysRevA.100.012126} {\bibfield  {journal} {\bibinfo
   {journal} {Phys. Rev. A}\ }\textbf {\bibinfo {volume} {100}},\ \bibinfo
  {pages} {012126} (\bibinfo {year} {2019})}\BibitemShut {NoStop}%
\bibitem [{\citenamefont {Cavina}\ \emph {et~al.}(2018)\citenamefont {Cavina},
  \citenamefont {Mari}, \citenamefont {Carlini},\ and\ \citenamefont
  {Giovannetti}}]{cavina2018}%
  \BibitemOpen
  \bibfield  {author} {\bibinfo {author} {\bibfnamefont {V.}~\bibnamefont
  {Cavina}}, \bibinfo {author} {\bibfnamefont {A.}~\bibnamefont {Mari}},
  \bibinfo {author} {\bibfnamefont {A.}~\bibnamefont {Carlini}},\ and\ \bibinfo
  {author} {\bibfnamefont {V.}~\bibnamefont {Giovannetti}},\ }\bibfield
  {title} {\bibinfo {title} {Optimal thermodynamic control in open quantum
  systems},\ }\href {https://doi.org/10.1103/PhysRevA.98.012139} {\bibfield
  {journal} {\bibinfo  {journal} {Phys. Rev. A}\ }\textbf {\bibinfo {volume}
  {98}},\ \bibinfo {pages} {012139} (\bibinfo {year} {2018})}\BibitemShut
  {NoStop}%
\bibitem [{\citenamefont {Suri}\ \emph {et~al.}(2018)\citenamefont {Suri},
  \citenamefont {Binder},\ and\ \citenamefont {Muralidharan}}]{suri2018}%
  \BibitemOpen
  \bibfield  {author} {\bibinfo {author} {\bibfnamefont {N.}~\bibnamefont
  {Suri}}, \bibinfo {author} {\bibfnamefont {F.~C.}\ \bibnamefont {Binder}},
  \bibinfo {author} {\bibfnamefont {B.}~\bibnamefont {Muralidharan}},\ and\ 
   \bibinfo {author} {\bibfnamefont {S.}~\bibnamefont {Vinjanampathy}},\ }\bibfield  {title} {\bibinfo {title}
  {Speeding up thermalisation via open quantum system variational
  optimisation},\ }\href {https://doi.org/10.1140/epjst/e2018-00125-6}
  {\bibfield  {journal} {\bibinfo  {journal} {Eur. Phys. J. Spec. Top.}\
  }\textbf {\bibinfo {volume} {227}},\ \bibinfo {pages} {203} (\bibinfo {year}
  {2018})}\BibitemShut {NoStop}%
  \bibitem [{\citenamefont {Menczel}\ \emph {et~al.}(2019)\citenamefont
  {Menczel}, \citenamefont {Pyh{\"a}ranta}, \citenamefont {Flindt},\ and\
  \citenamefont {Brandner}}]{menczel2019_prb}%
  \BibitemOpen
  \bibfield  {author} {\bibinfo {author} {\bibfnamefont {P.}~\bibnamefont
  {Menczel}}, \bibinfo {author} {\bibfnamefont {T.}~\bibnamefont
  {Pyh{\"a}ranta}}, \bibinfo {author} {\bibfnamefont {C.}~\bibnamefont
  {Flindt}},\ and\ \bibinfo {author} {\bibfnamefont {K.}~\bibnamefont
  {Brandner}},\ }\bibfield  {title} {\bibinfo {title} {Two-stroke optimization
  scheme for mesoscopic refrigerators},\ }\href
  {https://doi.org/10.1103/PhysRevB.99.224306} {\bibfield  {journal} {\bibinfo
  {journal} {Phys. Rev. B}\ }\textbf {\bibinfo {volume} {99}},\ \bibinfo
  {pages} {224306} (\bibinfo {year} {2019})}\BibitemShut {NoStop}%
\bibitem [{\citenamefont {Scully}\ \emph {et~al.}(2011)\citenamefont {Scully},
  \citenamefont {Chapin}, \citenamefont {Dorfman}, \citenamefont {Kim},\ and\
  \citenamefont {Svidzinsky}}]{scully2011}%
  \BibitemOpen
  \bibfield  {author} {\bibinfo {author} {\bibfnamefont {M.~O.}\ \bibnamefont
  {Scully}}, \bibinfo {author} {\bibfnamefont {K.~R.}\ \bibnamefont {Chapin}},
  \bibinfo {author} {\bibfnamefont {K.~E.}\ \bibnamefont {Dorfman}}, \bibinfo
  {author} {\bibfnamefont {M.~B.}\ \bibnamefont {Kim}},\ and\ \bibinfo {author}
  {\bibfnamefont {A.}~\bibnamefont {Svidzinsky}},\ }\bibfield  {title}
  {\bibinfo {title} {Quantum heat engine power can be increased by
  noise-induced coherence},\ }\href {https://doi.org/10.1073/pnas.1110234108}
  {\bibfield  {journal} {\bibinfo  {journal} {Proc. Natl. Acad. Sci. U.S.A.}\
  }\textbf {\bibinfo {volume} {108}},\ \bibinfo {pages} {15097} (\bibinfo
  {year} {2011})}\BibitemShut {NoStop}%
\bibitem [{\citenamefont {Uzdin}\ \emph {et~al.}(2015)\citenamefont {Uzdin},
  \citenamefont {Levy},\ and\ \citenamefont {Kosloff}}]{uzdin2015}%
  \BibitemOpen
  \bibfield  {author} {\bibinfo {author} {\bibfnamefont {R.}~\bibnamefont
  {Uzdin}}, \bibinfo {author} {\bibfnamefont {A.}~\bibnamefont {Levy}},\ and\
  \bibinfo {author} {\bibfnamefont {R.}~\bibnamefont {Kosloff}},\ }\bibfield
  {title} {\bibinfo {title} {Equivalence of quantum heat machines, and
  quantum-thermodynamic signatures},\ }\href
  {https://link.aps.org/doi/10.1103/PhysRevX.5.031044} {\bibfield  {journal}
  {\bibinfo  {journal} {Phys. Rev. X}\ }\textbf {\bibinfo {volume} {5}},\
  \bibinfo {pages} {031044} (\bibinfo {year} {2015})}\BibitemShut {NoStop}%
\bibitem [{\citenamefont {Jaramillo}\ \emph {et~al.}(2016)\citenamefont
  {Jaramillo}, \citenamefont {Beau},\ and\ \citenamefont {del
  Campo}}]{jaramillo2016}%
  \BibitemOpen
  \bibfield  {author} {\bibinfo {author} {\bibfnamefont {J.}~\bibnamefont
  {Jaramillo}}, \bibinfo {author} {\bibfnamefont {M.}~\bibnamefont {Beau}},\
  and\ \bibinfo {author} {\bibfnamefont {A.}~\bibnamefont {del Campo}},\
  }\bibfield  {title} {\bibinfo {title} {Quantum supremacy of many-particle
  thermal machines},\ }\href {https://doi.org/10.1088/1367-2630/18/7/075019}
  {\bibfield  {journal} {\bibinfo  {journal} {New J. Phys.}\ }\textbf {\bibinfo
  {volume} {18}},\ \bibinfo {pages} {075019} (\bibinfo {year}
  {2016})}\BibitemShut {NoStop}%
\bibitem [{\citenamefont {Brandner}\ \emph {et~al.}(2017)\citenamefont
  {Brandner}, \citenamefont {Bauer},\ and\ \citenamefont {U.}}]{brandner2017}%
  \BibitemOpen
  \bibfield  {author} {\bibinfo {author} {\bibfnamefont {K.}~\bibnamefont
  {Brandner}}, \bibinfo {author} {\bibfnamefont {M.}~\bibnamefont {Bauer}},\
  and\ \bibinfo {author} {\bibfnamefont {U.}~\bibnamefont {Seifert}},\ }\bibfield
  {title} {\bibinfo {title} {Universal coherence-induced power losses of
  quantum heat engines in linear response},\ }\href
  {https://doi.org/10.1103/PhysRevLett.119.170602} {\bibfield  {journal}
  {\bibinfo  {journal} {Phys. Rev. Lett.}\ }\textbf {\bibinfo {volume} {119}},\
  \bibinfo {pages} {170602} (\bibinfo {year} {2017})}\BibitemShut {NoStop}%
\bibitem [{\citenamefont {Kosloff}\ and\ \citenamefont
  {Feldmann}(2002)}]{kosloff2002}%
  \BibitemOpen
  \bibfield  {author} {\bibinfo {author} {\bibfnamefont {R.}~\bibnamefont
  {Kosloff}}\ and\ \bibinfo {author} {\bibfnamefont {T.}~\bibnamefont
  {Feldmann}},\ }\bibfield  {title} {\bibinfo {title} {Discrete four-stroke
  quantum heat engine exploring the origin of friction},\ }\href
  {https://doi.org/10.1103/PhysRevE.65.055102} {\bibfield  {journal} {\bibinfo
  {journal} {Phys. Rev. E}\ }\textbf {\bibinfo {volume} {65}},\ \bibinfo
  {pages} {055102} (\bibinfo {year} {2002})}\BibitemShut {NoStop}%
\bibitem [{\citenamefont {Sutton}\ and\ \citenamefont
  {Barto}(2018)}]{sutton2018}%
  \BibitemOpen
  \bibfield  {author} {\bibinfo {author} {\bibfnamefont {R.~S.}\ \bibnamefont
  {Sutton}}\ and\ \bibinfo {author} {\bibfnamefont {A.~G.}\ \bibnamefont
  {Barto}},\ }\href@noop {} {\emph {\bibinfo {title} {Reinforcement learning:
  An introduction}}}\ (\bibinfo  {publisher} {MIT press},\ \bibinfo {year}
  {2018})\BibitemShut {NoStop}%
\bibitem [{\citenamefont {Haarnoja}\ \emph
  {et~al.}(2018{\natexlab{a}})\citenamefont {Haarnoja}, \citenamefont {Zhou},
  \citenamefont {Abbeel},\ and\ \citenamefont {Levine}}]{haarnoja2018_pmlr}%
  \BibitemOpen
  \bibfield  {author} {\bibinfo {author} {\bibfnamefont {T.}~\bibnamefont
  {Haarnoja}}, \bibinfo {author} {\bibfnamefont {A.}~\bibnamefont {Zhou}},
  \bibinfo {author} {\bibfnamefont {P.}~\bibnamefont {Abbeel}},\ and\ \bibinfo
  {author} {\bibfnamefont {S.}~\bibnamefont {Levine}},\ }\bibfield  {title}
  {\bibinfo {title} {Soft actor-critic: Off-policy maximum entropy deep
  reinforcement learning with a stochastic actor},\ }in\ \href
  {http://proceedings.mlr.press/v80/haarnoja18b.html} {\emph {\bibinfo
  {booktitle} {International Conference on Machine Learning}}},\ Vol.~\bibinfo
  {volume} {80}\ (\bibinfo {organization} {PMLR},\ \bibinfo {year} {2018})\ p.\
  \bibinfo {pages} {1861}\BibitemShut {NoStop}%
\bibitem [{\citenamefont {Haarnoja}\ \emph
  {et~al.}(2018{\natexlab{b}})\citenamefont {Haarnoja}, \citenamefont {Zhou},
  \citenamefont {Hartikainen}, \citenamefont {Tucker}, \citenamefont {Ha},
  \citenamefont {Tan}, \citenamefont {Kumar}, \citenamefont {Zhu},
  \citenamefont {Gupta}, \citenamefont {Abbeel},\ and\ \citenamefont
  {Levine}}]{haarnoja2018_arxiv_sac}%
  \BibitemOpen
  \bibfield  {author} {\bibinfo {author} {\bibfnamefont {T.}~\bibnamefont
  {Haarnoja}}  \textit{et al.,} }\bibfield  {title} {\bibinfo
  {title} {Soft actor-critic algorithms and applications}.\ } Preprint at \Eprint
  {https://arxiv.org/abs/1812.05905} {https://arxiv.org/abs/1812.05905}  (\bibinfo {year}
  {2018}{\natexlab{b}})\BibitemShut {NoStop}%
\bibitem [{\citenamefont {Christodoulou}(2019)}]{christodoulou2019}%
  \BibitemOpen
  \bibfield  {author} {\bibinfo {author} {\bibfnamefont {P.}~\bibnamefont
  {Christodoulou}},\ }\bibfield  {title} {\bibinfo {title} {Soft actor-critic
  for discrete action settings}.\ } Preprint at \Eprint
  {https://arxiv.org/abs/1910.07207} {https://arxiv.org/abs/1910.07207}  (\bibinfo {year}
  {2019})\BibitemShut {NoStop}%
\bibitem [{\citenamefont {Delalleau}\ \emph {et~al.}(2019)\citenamefont
  {Delalleau}, \citenamefont {Peter}, \citenamefont {Alonso},\ and\
  \citenamefont {Logut}}]{delalleau2019}%
  \BibitemOpen
  \bibfield  {author} {\bibinfo {author} {\bibfnamefont {O.}~\bibnamefont
  {Delalleau}}, \bibinfo {author} {\bibfnamefont {M.}~\bibnamefont {Peter}},
  \bibinfo {author} {\bibfnamefont {E.}~\bibnamefont {Alonso}},\ and\ \bibinfo
  {author} {\bibfnamefont {A.}~\bibnamefont {Logut}},\ }\bibfield  {title}
  {\bibinfo {title} {Discrete and continuous action representation for
  practical rl in video games}.\ } Preprint at \Eprint
  {https://arxiv.org/abs/1912.11077} {https://arxiv.org/abs/1912.11077}  (\bibinfo {year}
  {2019})\BibitemShut {NoStop}%
\bibitem [{\citenamefont {Mnih}\ \emph {et~al.}(2015)\citenamefont {Mnih},
  \citenamefont {Kavukcuoglu}, \citenamefont {Silver}, \citenamefont {Rusu},
  \citenamefont {Veness}, \citenamefont {Bellemare}, \citenamefont {Graves},
  \citenamefont {Riedmiller}, \citenamefont {Fidjeland}, \citenamefont
  {Ostrovski} \emph {et~al.}}]{mnih2015}%
  \BibitemOpen
  \bibfield  {author} {\bibinfo {author} {\bibfnamefont {V.}~\bibnamefont
  {Mnih}}  \textit{et al.,} }\bibfield  {title} {\bibinfo {title}
  {Human-level control through deep reinforcement learning},\ }\href
  {https://doi.org/10.1038/nature14236} {\bibfield  {journal} {\bibinfo
  {journal} {Nature}\ }\textbf {\bibinfo {volume} {518}},\ \bibinfo {pages}
  {529} (\bibinfo {year} {2015})}\BibitemShut {NoStop}%
\bibitem [{\citenamefont {Vinyals}\ \emph {et~al.}(2019)\citenamefont
  {Vinyals}, \citenamefont {Babuschkin}, \citenamefont {Czarnecki},
  \citenamefont {Mathieu}, \citenamefont {Dudzik}, \citenamefont {Chung},
  \citenamefont {Choi}, \citenamefont {Powell}, \citenamefont {Ewalds},
  \citenamefont {Georgiev} \emph {et~al.}}]{vinyals2019}%
  \BibitemOpen
  \bibfield  {author} {\bibinfo {author} {\bibfnamefont {O.}~\bibnamefont
  {Vinyals}}  \textit{et al.,} }\bibfield  {title} {\bibinfo {title} {Grandmaster level in
  starcraft ii using multi-agent reinforcement learning},\ }\href
  {https://doi.org/10.1038/s41586-019-1724-z} {\bibfield  {journal} {\bibinfo
  {journal} {Nature}\ }\textbf {\bibinfo {volume} {575}},\ \bibinfo {pages}
  {350} (\bibinfo {year} {2019})}\BibitemShut {NoStop}%
\bibitem [{\citenamefont {Silver}\ \emph {et~al.}(2017)\citenamefont {Silver},
  \citenamefont {Schrittwieser}, \citenamefont {Simonyan}, \citenamefont
  {Antonoglou}, \citenamefont {Huang}, \citenamefont {Guez}, \citenamefont
  {Hubert}, \citenamefont {Baker}, \citenamefont {Lai}, \citenamefont {Bolton}
  \emph {et~al.}}]{silver2017}%
  \BibitemOpen
  \bibfield  {author} {\bibinfo {author} {\bibfnamefont {D.}~\bibnamefont
  {Silver}}  \textit{et al.,} }\bibfield  {title} {\bibinfo {title} {Mastering
  the game of go without human knowledge},\ }\href
  {https://doi.org/10.1038/nature24270} {\bibfield  {journal} {\bibinfo
  {journal} {Nature}\ }\textbf {\bibinfo {volume} {550}},\ \bibinfo {pages}
  {354} (\bibinfo {year} {2017})}\BibitemShut {NoStop}%
\bibitem [{\citenamefont {Haarnoja}\ \emph
  {et~al.}(2018{\natexlab{c}})\citenamefont {Haarnoja}, \citenamefont {Ha},
  \citenamefont {Zhou}, \citenamefont {Tan}, \citenamefont {Tucker},\ and\
  \citenamefont {Levine}}]{haarnoja2018_arxiv_walk}%
  \BibitemOpen
  \bibfield  {author} {\bibinfo {author} {\bibfnamefont {T.}~\bibnamefont
  {Haarnoja}}  \textit{et al.,} }\bibfield  {title} {\bibinfo {title} {Learning
  to walk via deep reinforcement learning}.\ } Preprint at \Eprint
  {https://arxiv.org/abs/1812.11103} {https://arxiv.org/abs/1812.11103}  (\bibinfo {year}
  {2018}{\natexlab{c}})\BibitemShut {NoStop}%
\bibitem [{\citenamefont {Bukov}\ \emph {et~al.}(2018)\citenamefont {Bukov},
  \citenamefont {Day}, \citenamefont {Sels}, \citenamefont {Weinberg},
  \citenamefont {Polkovnikov},\ and\ \citenamefont {Mehta}}]{bukov2018}%
  \BibitemOpen
  \bibfield  {author} {\bibinfo {author} {\bibfnamefont {M.}~\bibnamefont
  {Bukov}} \textit{et al.,} }\bibfield  {title} {\bibinfo
  {title} {Reinforcement learning in different phases of quantum control},\
  }\href {https://doi.org/10.1103/PhysRevX.8.031086} {\bibfield  {journal}
  {\bibinfo  {journal} {Phys. Rev. X}\ }\textbf {\bibinfo {volume} {8}},\
  \bibinfo {pages} {031086} (\bibinfo {year} {2018})}\BibitemShut {NoStop}%
\bibitem [{\citenamefont {An}\ and\ \citenamefont {Zhou}(2019)}]{an2019}%
  \BibitemOpen
  \bibfield  {author} {\bibinfo {author} {\bibfnamefont {Z.}~\bibnamefont
  {An}}\ and\ \bibinfo {author} {\bibfnamefont {D.}~\bibnamefont {Zhou}},\
  }\bibfield  {title} {\bibinfo {title} {Deep reinforcement learning for
  quantum gate control},\ }\href {https://doi.org/10.1209/0295-5075/126/60002}
  {\bibfield  {journal} {\bibinfo  {journal} {EPL}\ }\textbf {\bibinfo {volume}
  {126}},\ \bibinfo {pages} {60002} (\bibinfo {year} {2019})}\BibitemShut
  {NoStop}%
\bibitem [{\citenamefont {Dalgaard}\ \emph {et~al.}(2020)\citenamefont
  {Dalgaard}, \citenamefont {Motzoi}, \citenamefont {S{\o}rensen},\ and\
  \citenamefont {Sherson}}]{dalgaard2020}%
  \BibitemOpen
  \bibfield  {author} {\bibinfo {author} {\bibfnamefont {M.}~\bibnamefont
  {Dalgaard}}, \bibinfo {author} {\bibfnamefont {F.}~\bibnamefont {Motzoi}},
  \bibinfo {author} {\bibfnamefont {J.~J.}\ \bibnamefont {S{\o}rensen}},\ and\
  \bibinfo {author} {\bibfnamefont {J.}~\bibnamefont {Sherson}},\ }\bibfield
  {title} {\bibinfo {title} {Global optimization of quantum dynamics with
  alphazero deep exploration},\ }\href
  {https://doi.org/10.1038/s41534-019-0241-0} {\bibfield  {journal} {\bibinfo
  {journal} {NPJ Quantum Inf.}\ }\textbf {\bibinfo {volume} {6}},\ \bibinfo
  {pages} {6} (\bibinfo {year} {2020})}\BibitemShut {NoStop}%
\bibitem [{\citenamefont {Mackeprang}\ \emph {et~al.}(2020)\citenamefont
  {Mackeprang}, \citenamefont {Dasari},\ and\ \citenamefont
  {Wrachtrup}}]{mackeprang2020}%
  \BibitemOpen
  \bibfield  {author} {\bibinfo {author} {\bibfnamefont {J.}~\bibnamefont
  {Mackeprang}}, \bibinfo {author} {\bibfnamefont {D.~B.~R.}\ \bibnamefont
  {Dasari}},\ and\ \bibinfo {author} {\bibfnamefont {J.}~\bibnamefont
  {Wrachtrup}},\ }\bibfield  {title} {\bibinfo {title} {A reinforcement
  learning approach for quantum state engineering},\ }\href
  {https://doi.org/10.1007/s42484-020-00016-8} {\bibfield  {journal} {\bibinfo
  {journal} {Quantum Mach. Intell.}\ }\textbf {\bibinfo {volume} {2}},\
  \bibinfo {pages} {5} (\bibinfo {year} {2020})}\BibitemShut {NoStop}%
\bibitem [{\citenamefont {Niu}\ \emph {et~al.}(2019)\citenamefont {Niu},
  \citenamefont {Boixo}, \citenamefont {Smelyanskiy},\ and\ \citenamefont
  {Neven}}]{niu2019}%
  \BibitemOpen
  \bibfield  {author} {\bibinfo {author} {\bibfnamefont {M.~Y.}\ \bibnamefont
  {Niu}}, \bibinfo {author} {\bibfnamefont {S.}~\bibnamefont {Boixo}}, \bibinfo
  {author} {\bibfnamefont {V.~N.}\ \bibnamefont {Smelyanskiy}},\ and\ \bibinfo
  {author} {\bibfnamefont {H.}~\bibnamefont {Neven}},\ }\bibfield  {title}
  {\bibinfo {title} {Universal quantum control through deep reinforcement
  learning},\ }\href {https://doi.org/10.1038/s41534-019-0141-3} {\bibfield
  {journal} {\bibinfo  {journal} {NPJ Quantum Inf.}\ }\textbf {\bibinfo
  {volume} {5}},\ \bibinfo {pages} {33} (\bibinfo {year} {2019})}\BibitemShut
  {NoStop}%
\bibitem [{\citenamefont {Zhang}\ \emph {et~al.}(2019)\citenamefont {Zhang},
  \citenamefont {Wei}, \citenamefont {Asad}, \citenamefont {Yang},\ and\
  \citenamefont {Wang}}]{zhang2019}%
  \BibitemOpen
  \bibfield  {author} {\bibinfo {author} {\bibfnamefont {X.-M.}\ \bibnamefont
  {Zhang}}, \bibinfo {author} {\bibfnamefont {Z.}~\bibnamefont {Wei}}, \bibinfo
  {author} {\bibfnamefont {R.}~\bibnamefont {Asad}}, \bibinfo {author}
  {\bibfnamefont {X.-C.}\ \bibnamefont {Yang}},\ and\ \bibinfo {author}
  {\bibfnamefont {X.}~\bibnamefont {Wang}},\ }\bibfield  {title} {\bibinfo
  {title} {When does reinforcement learning stand out in quantum control? a
  comparative study on state preparation},\ }\href
  {https://doi.org/10.1038/s41534-019-0201-8} {\bibfield  {journal} {\bibinfo
  {journal} {NPJ Quantum Inf.}\ }\textbf {\bibinfo {volume} {5}},\ \bibinfo
  {pages} {85} (\bibinfo {year} {2019})}\BibitemShut {NoStop}%
\bibitem [{\citenamefont {Sgroi}\ \emph {et~al.}(2021)\citenamefont {Sgroi},
  \citenamefont {Palma},\ and\ \citenamefont {Paternostro}}]{sgroi2021}%
  \BibitemOpen
  \bibfield  {author} {\bibinfo {author} {\bibfnamefont {P.}~\bibnamefont
  {Sgroi}}, \bibinfo {author} {\bibfnamefont {G.~M.}\ \bibnamefont {Palma}},\
  and\ \bibinfo {author} {\bibfnamefont {M.}~\bibnamefont {Paternostro}},\
  }\bibfield  {title} {\bibinfo {title} {Reinforcement learning approach to
  nonequilibrium quantum thermodynamics},\ }\href
  {https://doi.org/10.1103/PhysRevLett.126.020601} {\bibfield  {journal}
  {\bibinfo  {journal} {Phys. Rev. Lett.}\ }\textbf {\bibinfo {volume} {126}},\
  \bibinfo {pages} {020601} (\bibinfo {year} {2021})}\BibitemShut {NoStop}%
\bibitem{sweke2021}%
  \BibitemOpen
  \bibfield  {author} {
  \bibinfo {author} {\bibfnamefont {R.}~\bibnamefont {Sweke}}, 
  \bibinfo {author} {\bibfnamefont {M.~S.}\ \bibnamefont {Kesselring}},
  \bibinfo {author} {\bibfnamefont {E.~P.~L.}\ \bibnamefont {van Nieuwenburg}},
  \ and\ \bibinfo {author} {\bibfnamefont {J.}~\bibnamefont {Eisert}},\
  }\bibfield  {title} {\bibinfo {title} {Reinforcement learning decoders for fault-tolerant quantum computation},\ }\href
  {https://doi.org/10.1088/2632-2153/abc609} {\bibfield  {journal}
  {\bibinfo  {journal} { Mach. Learn.: Sci. Technol.}\ }\textbf {\bibinfo {volume} {2}},\
  \bibinfo {pages} {025005} (\bibinfo {year} {2020})}\BibitemShut {NoStop}%
\bibitem{luiz2021}%
  \BibitemOpen
  \bibfield  {author} {
  \bibinfo {author} {\bibfnamefont {F.~S.}\ \bibnamefont {Luiz}},  
  \bibinfo {author} {\bibfnamefont {A.}~\bibnamefont {de Oliveira Junior}},\
  \bibinfo {author} {\bibfnamefont {F.~F.}~\bibnamefont {Fanchini}},\ and\
  \bibinfo {author} {\bibfnamefont {G.~T.}~\bibnamefont {Landi}},\
  }
  \bibfield  {title} {\bibinfo {title} {Machine classification for probe based quantum thermometry}.\ } Preprint at \Eprint
  {https://arxiv.org/abs/2107.04555}
  {https://arxiv.org/abs/2107.04555}  (\bibinfo {year} {2021})\BibitemShut
  {NoStop}%
\bibitem [{\citenamefont {Erdman}\ \emph {et~al.}(2019)\citenamefont {Erdman},
  \citenamefont {Cavina}, \citenamefont {Fazio}, \citenamefont {Taddei},\ and\
  \citenamefont {Giovannetti}}]{erdman2019_njp}%
  \BibitemOpen
  \bibfield  {author} {\bibinfo {author} {\bibfnamefont {P.~A.}\ \bibnamefont
  {Erdman}}, \bibinfo {author} {\bibfnamefont {V.}~\bibnamefont {Cavina}},
  \bibinfo {author} {\bibfnamefont {R.}~\bibnamefont {Fazio}}, \bibinfo
  {author} {\bibfnamefont {F.}~\bibnamefont {Taddei}},\ and\ \bibinfo {author}
  {\bibfnamefont {V.}~\bibnamefont {Giovannetti}},\ }\bibfield  {title}
  {\bibinfo {title} {Maximum power and corresponding efficiency for two-level
  heat engines and refrigerators: optimality of fast cycles},\ }\href
  {https://doi.org/10.1088/1367-2630/ab4dca} {\bibfield  {journal} {\bibinfo
  {journal} {New J. Phys.}\ }\textbf {\bibinfo {volume} {21}},\ \bibinfo
  {pages} {103049} (\bibinfo {year} {2019})}\BibitemShut {NoStop}%
\bibitem [{\citenamefont {Lekscha}\ \emph {et~al.}(2018)\citenamefont
  {Lekscha}, \citenamefont {Wilming}, \citenamefont {Eisert},\ and\
  \citenamefont {Gallego}}]{lekscha2018}%
  \BibitemOpen
  \bibfield  {author} {\bibinfo {author} {\bibfnamefont {J.}~\bibnamefont
  {Lekscha}}, \bibinfo {author} {\bibfnamefont {H.}~\bibnamefont {Wilming}},
  \bibinfo {author} {\bibfnamefont {J.}~\bibnamefont {Eisert}},\ and\ \bibinfo
  {author} {\bibfnamefont {R.}~\bibnamefont {Gallego}},\ }\bibfield  {title}
  {\bibinfo {title} {Quantum thermodynamics with local control},\ }\href
  {https://doi.org/10.1103/PhysRevE.97.022142} {\bibfield  {journal} {\bibinfo
  {journal} {Phys. Rev. E}\ }\textbf {\bibinfo {volume} {97}},\ \bibinfo
  {pages} {022142} (\bibinfo {year} {2018})}\BibitemShut {NoStop}%
\bibitem [{\citenamefont {Gorini}\ \emph {et~al.}(1976)\citenamefont {Gorini},
  \citenamefont {Kossakowski},\ and\ \citenamefont {Sudarshan}}]{gorini1976}%
  \BibitemOpen
  \bibfield  {author} {\bibinfo {author} {\bibfnamefont {V.}~\bibnamefont
  {Gorini}}, \bibinfo {author} {\bibfnamefont {A.}~\bibnamefont
  {Kossakowski}},\ and\ \bibinfo {author} {\bibfnamefont {E.~C.~G.}\
  \bibnamefont {Sudarshan}},\ }\bibfield  {title} {\bibinfo {title} {Completely
  positive dynamical semigroups of {N}‐level systems},\ }\href
  {https://doi.org/10.1063/1.522979} {\bibfield  {journal} {\bibinfo  {journal}
  {J. Math. Phys.}\ }\textbf {\bibinfo {volume} {17}},\ \bibinfo {pages} {821}
  (\bibinfo {year} {1976})}\BibitemShut {NoStop}%
\bibitem [{\citenamefont {Lindblad}(1976)}]{lindblad1976}%
  \BibitemOpen
  \bibfield  {author} {\bibinfo {author} {\bibfnamefont {G.}~\bibnamefont
  {Lindblad}},\ }\bibfield  {title} {\bibinfo {title} {On the generators of
  quantum dynamical semigroups},\ }\href {https://doi.org/10.1007/BF01608499}
  {\bibfield  {journal} {\bibinfo  {journal} {Commun. Math. Phys}\ }\textbf
  {\bibinfo {volume} {48}},\ \bibinfo {pages} {119} (\bibinfo {year}
  {1976})}\BibitemShut {NoStop}%
\bibitem [{\citenamefont {Breuer}\ and\ \citenamefont
  {Petruccione}(2002)}]{breuer2002}%
  \BibitemOpen
  \bibfield  {author} {\bibinfo {author} {\bibfnamefont {H.}~\bibnamefont
  {Breuer}}\ and\ \bibinfo {author} {\bibfnamefont {F.}~\bibnamefont
  {Petruccione}},\ }\href@noop {} {\emph {\bibinfo {title} {The theory of open
  quantum systems}}}\ (\bibinfo  {publisher} {Oxford University Press},\
  \bibinfo {year} {2002})\BibitemShut {NoStop}%
\bibitem [{\citenamefont {Yamaguchi}\ \emph {et~al.}(2017)\citenamefont
  {Yamaguchi}, \citenamefont {Yuge},\ and\ \citenamefont
  {Ogawa}}]{yamaguchi2017}%
  \BibitemOpen
  \bibfield  {author} {\bibinfo {author} {\bibfnamefont {M.}~\bibnamefont
  {Yamaguchi}}, \bibinfo {author} {\bibfnamefont {T.}~\bibnamefont {Yuge}},\
  and\ \bibinfo {author} {\bibfnamefont {T.}~\bibnamefont {Ogawa}},\ }\bibfield
   {title} {\bibinfo {title} {Markovian quantum master equation beyond
  adiabatic regime},\ }\href {https://doi.org/10.1103/PhysRevE.95.012136}
  {\bibfield  {journal} {\bibinfo  {journal} {Phys. Rev. E}\ }\textbf {\bibinfo
  {volume} {95}},\ \bibinfo {pages} {012136} (\bibinfo {year}
  {2017})}\BibitemShut {NoStop}%
\bibitem [{\citenamefont {Kingma}\ and\ \citenamefont {Ba}(2014)}]{kingma2014}%
  \BibitemOpen
  \bibfield  {author} {\bibinfo {author} {\bibfnamefont {D.~P.}\ \bibnamefont
  {Kingma}}\ and\ \bibinfo {author} {\bibfnamefont {J.}~\bibnamefont {Ba}},\
  }\bibfield  {title} {\bibinfo {title} {Adam: A method for stochastic
  optimization}.\ } Preprint at \Eprint
  {https://arxiv.org/abs/1412.6980}
  {https://arxiv.org/abs/1412.6980}  (\bibinfo {year} {2014})\BibitemShut
  {NoStop}%
\bibitem [{\citenamefont {Curzon}\ and\ \citenamefont
  {Ahlborn}(1975)}]{curzon1975}%
  \BibitemOpen
  \bibfield  {author} {\bibinfo {author} {\bibfnamefont {F.}~\bibnamefont
  {Curzon}}\ and\ \bibinfo {author} {\bibfnamefont {B.}~\bibnamefont
  {Ahlborn}},\ }\bibfield  {title} {\bibinfo {title} {Efficiency of a carnot
  engine at maximum power output},\ }\href {https://doi.org/10.1119/1.10023}
  {\bibfield  {journal} {\bibinfo  {journal} {Am. J. Phys.}\ }\textbf {\bibinfo
  {volume} {43}},\ \bibinfo {pages} {22} (\bibinfo {year} {1975})}\BibitemShut
  {NoStop}%
\bibitem [{\citenamefont {Schmiedl}\ and\ \citenamefont
  {Seifert}(2007)}]{schmiedl2007}%
  \BibitemOpen
  \bibfield  {author} {\bibinfo {author} {\bibfnamefont {T.}~\bibnamefont
  {Schmiedl}}\ and\ \bibinfo {author} {\bibfnamefont {U.}~\bibnamefont
  {Seifert}},\ }\bibfield  {title} {\bibinfo {title} {Efficiency at maximum
  power: An analytically solvable model for stochastic heat engines},\ }\href
  {https://doi.org/10.1209/0295-5075/81/20003} {\bibfield  {journal} {\bibinfo
  {journal} {Europhys. Lett.}\ }\textbf {\bibinfo {volume} {81}},\ \bibinfo
  {pages} {20003} (\bibinfo {year} {2007})}\BibitemShut {NoStop}%
\bibitem [{\citenamefont {den Broeck}(2005)}]{broeck2005}%
  \BibitemOpen
  \bibfield  {author} {\bibinfo {author} {\bibfnamefont {C.~V.}\ \bibnamefont
  {den Broeck}},\ }\bibfield  {title} {\bibinfo {title} {Thermodynamic
  efficiency at maximum power},\ }\href
  {https://doi.org/10.1103/PhysRevLett.95.190602} {\bibfield  {journal}
  {\bibinfo  {journal} {Phys. Rev. Lett.}\ }\textbf {\bibinfo {volume} {95}},\
  \bibinfo {pages} {190602} (\bibinfo {year} {2005})}\BibitemShut {NoStop}%
\bibitem [{\citenamefont {Baumgratz}\ \emph {et~al.}(2014)\citenamefont
  {Baumgratz}, \citenamefont {Cramer},\ and\ \citenamefont
  {Plenio}}]{baumgratz2014}%
  \BibitemOpen
  \bibfield  {author} {\bibinfo {author} {\bibfnamefont {T.}~\bibnamefont
  {Baumgratz}}, \bibinfo {author} {\bibfnamefont {M.}~\bibnamefont {Cramer}},\
  and\ \bibinfo {author} {\bibfnamefont {M.~B.}\ \bibnamefont {Plenio}},\
  }\bibfield  {title} {\bibinfo {title} {Quantifying coherence},\ }\href
  {https://doi.org/10.1103/PhysRevLett.113.140401} {\bibfield  {journal}
  {\bibinfo  {journal} {Phys. Rev. Lett.}\ }\textbf {\bibinfo {volume} {113}},\
  \bibinfo {pages} {140401} (\bibinfo {year} {2014})}\BibitemShut {NoStop}%
\bibitem [{\citenamefont {Gallego}\ \emph {et~al.}(2014)\citenamefont
  {Gallego}, \citenamefont {Riera},\ and\ \citenamefont
  {Eisert}}]{gallego2014}%
  \BibitemOpen
  \bibfield  {author} {\bibinfo {author} {\bibfnamefont {R.}~\bibnamefont
  {Gallego}}, \bibinfo {author} {\bibfnamefont {A.}~\bibnamefont {Riera}},\
  and\ \bibinfo {author} {\bibfnamefont {J.}~\bibnamefont {Eisert}},\
  }\bibfield  {title} {\bibinfo {title} {Thermal machines beyond the weak
  coupling regime},\ }\href {https://doi.org/10.1088/1367-2630/16/12/125009}
  {\bibfield  {journal} {\bibinfo  {journal} {New J. Phys.}\ }\textbf {\bibinfo
  {volume} {16}},\ \bibinfo {pages} {125009} (\bibinfo {year}
  {2014})}\BibitemShut {NoStop}%
\bibitem [{\citenamefont {Gelbwaser-Klimovsky}\ and\ \citenamefont
  {Aspuru-Guzik}(2015)}]{gelbwaser2015}%
  \BibitemOpen
  \bibfield  {author} {\bibinfo {author} {\bibfnamefont {D.}~\bibnamefont
  {Gelbwaser-Klimovsky}}\ and\ \bibinfo {author} {\bibfnamefont
  {A.}~\bibnamefont {Aspuru-Guzik}},\ }\bibfield  {title} {\bibinfo {title}
  {Strongly coupled quantum heat machines},\ }\href
  {https://doi.org/10.1021/acs.jpclett.5b01404} {\bibfield  {journal} {\bibinfo
   {journal} {J. Phys. Chem. Lett.}\ }\textbf {\bibinfo {volume} {6}},\
  \bibinfo {pages} {3477} (\bibinfo {year} {2015})}\BibitemShut {NoStop}%
\bibitem [{\citenamefont {Perarnau-Llobet}\ \emph {et~al.}(2018)\citenamefont
  {Perarnau-Llobet}, \citenamefont {Wilming}, \citenamefont {Riera},
  \citenamefont {Gallego},\ and\ \citenamefont {Eisert}}]{perarnau2018}%
  \BibitemOpen
  \bibfield  {author} {\bibinfo {author} {\bibfnamefont {M.}~\bibnamefont
  {Perarnau-Llobet}}, \bibinfo {author} {\bibfnamefont {H.}~\bibnamefont
  {Wilming}}, \bibinfo {author} {\bibfnamefont {A.}~\bibnamefont {Riera}},
  \bibinfo {author} {\bibfnamefont {R.}~\bibnamefont {Gallego}},\ and\ \bibinfo
  {author} {\bibfnamefont {J.}~\bibnamefont {Eisert}},\ }\bibfield  {title}
  {\bibinfo {title} {Strong coupling corrections in quantum thermodynamics},\
  }\href {https://doi.org/10.1103/PhysRevLett.120.120602} {\bibfield  {journal}
  {\bibinfo  {journal} {Phys. Rev. Lett.}\ }\textbf {\bibinfo {volume} {120}},\
  \bibinfo {pages} {120602} (\bibinfo {year} {2018})}\BibitemShut {NoStop}%
\bibitem [{\citenamefont {Dann}\ \emph {et~al.}(2018)\citenamefont {Dann},
  \citenamefont {Levy},\ and\ \citenamefont {Kosloff}}]{dann2018}%
  \BibitemOpen
  \bibfield  {author} {\bibinfo {author} {\bibfnamefont {R.}~\bibnamefont
  {Dann}}, \bibinfo {author} {\bibfnamefont {A.}~\bibnamefont {Levy}},\ and\
  \bibinfo {author} {\bibfnamefont {R.}~\bibnamefont {Kosloff}},\ }\bibfield
  {title} {\bibinfo {title} {Time-dependent markovian quantum master
  equation},\ }\href {https://doi.org/10.1103/PhysRevA.98.052129} {\bibfield
  {journal} {\bibinfo  {journal} {Phys. Rev. A}\ }\textbf {\bibinfo {volume}
  {98}},\ \bibinfo {pages} {052129} (\bibinfo {year} {2018})}\BibitemShut
  {NoStop}%
\bibitem [{\citenamefont {Beenakker}(1991)}]{beenakker1991}%
  \BibitemOpen
  \bibfield  {author} {\bibinfo {author} {\bibfnamefont {C.~W.~J.}\
  \bibnamefont {Beenakker}},\ }\bibfield  {title} {\bibinfo {title} {Theory of
  coulomb-blockade oscillations in the conductance of a quantum dot},\ }\href
  {https://doi.org/10.1103/PhysRevB.44.1646} {\bibfield  {journal} {\bibinfo
  {journal} {Phys. Rev. B}\ }\textbf {\bibinfo {volume} {44}},\ \bibinfo
  {pages} {1646} (\bibinfo {year} {1991})}\BibitemShut {NoStop}%
\bibitem [{\citenamefont {Esposito}\ \emph {et~al.}(2009)\citenamefont
  {Esposito}, \citenamefont {Lindenberg},\ and\ \citenamefont {den
  Broeck}}]{esposito2009}%
  \BibitemOpen
  \bibfield  {author} {\bibinfo {author} {\bibfnamefont {M.}~\bibnamefont
  {Esposito}}, \bibinfo {author} {\bibfnamefont {K.}~\bibnamefont
  {Lindenberg}},\ and\ \bibinfo {author} {\bibfnamefont {C.~V.}\ \bibnamefont
  {den Broeck}},\ }\bibfield  {title} {\bibinfo {title} {Thermoelectric
  efficiency at maximum power in a quantum dot},\ }\href
  {https://doi.org/10.1209/0295-5075/85/60010} {\bibfield  {journal} {\bibinfo
  {journal} {Eurphys. Lett.}\ }\textbf {\bibinfo {volume} {85}},\ \bibinfo
  {pages} {60010} (\bibinfo {year} {2009})}\BibitemShut {NoStop}%
\bibitem [{\citenamefont {Nazarov}\ and\ \citenamefont
  {Banter}(2009)}]{nazarov2009}%
  \BibitemOpen
  \bibfield  {author} {\bibinfo {author} {\bibfnamefont {Y.~V.}\ \bibnamefont
  {Nazarov}}\ and\ \bibinfo {author} {\bibfnamefont {Y.~M.}\ \bibnamefont
  {Banter}},\ }\href@noop {} {\emph {\bibinfo {title} {Quantum Transport}}}\
  (\bibinfo  {publisher} {Cambridge, New York},\ \bibinfo {year}
  {2009})\BibitemShut {NoStop}%
\bibitem [{\citenamefont {Erdman}\ \emph {et~al.}(2017)\citenamefont {Erdman},
  \citenamefont {Mazza}, \citenamefont {Bosisio}, \citenamefont {Benenti},
  \citenamefont {Fazio},\ and\ \citenamefont {Taddei}}]{erdman2017}%
  \BibitemOpen
  \bibfield  {author} {\bibinfo {author} {\bibfnamefont {P.~A.}\ \bibnamefont
  {Erdman}}  \textit{et al.,} }\bibfield  {title} {\bibinfo
  {title} {Thermoelectric properties of an interacting quantum dot based heat
  engine},\ }\href {https://doi.org/10.1103/PhysRevB.95.245432} {\bibfield
  {journal} {\bibinfo  {journal} {Phys. Rev. B}\ }\textbf {\bibinfo {volume}
  {95}},\ \bibinfo {pages} {245432} (\bibinfo {year} {2017})}\BibitemShut
  {NoStop}%
\bibitem [{\citenamefont {Achiam}(2018)}]{spinningup2018}%
  \BibitemOpen
  \bibfield  {author} {\bibinfo {author} {\bibfnamefont {J.}~\bibnamefont
  {Achiam}},\ }\bibfield  {title} {\bibinfo {title} {{Spinning Up in Deep
  Reinforcement Learning}},\ }\href
  {https://github.com/openai/spinningup}
  {https://github.com/openai/spinningup}  (\bibinfo {year} {2018})\BibitemShut
  {NoStop}%
\end{thebibliography}

%

\end{document}